%% file: 3mat-acgd-revised.tex
\documentclass[authoryear,11pt]{elsarticle}

\usepackage[cp1250]{inputenc}
\usepackage{graphics}
\usepackage{amssymb}
\usepackage{amsthm}
\usepackage{amsmath}
\usepackage{amsfonts}
\usepackage{amssymb}
\usepackage{here}

\newcommand{\cA}{\mathsf{A}}
\newcommand{\cB}{\mathsf{B}}
\newcommand{\cC}{\mathsf{C}}
\newcommand{\cD}{\mathsf{D}}
\newcommand{\cE}{\mathsf{E}}
\newcommand{\mE}{\mathbb{E}}
\newcommand{\mR}{\mathbb{R}}
\newcommand{\mT}{\mathbb{T}}
\newcommand{\ba}{\begin{array}}
\newcommand{\ea}{\end{array}}
\newcommand{\bc}{\begin{cases}}
\newcommand{\ec}{\end{cases}}
\newcommand{\be}{\begin{equation}}
\newcommand{\ee}{\end{equation}}
\newcommand{\bpm}{\begin{pmatrix}}
\newcommand{\epm}{\end{pmatrix}}
\newcommand{\2}{^{\,2}}
\newcommand{\mrm}{\mathrm}
\newcommand{\disp}{\displaystyle}

\journal{arxiv.org}

\begin{document}

\begin{frontmatter}

\title{Three-phase plane composites of minimal elastic stress energy: High-porosity structures}

\author[ac]{Andrej Cherkaev\corref{cor1}}
\ead{cherk@math.utah.edu}
\address[ac]{Department of Mathematics, University of Utah, Salt Lake City, UT, USA.}
\cortext[cor1]{Corresponding author. Address: Department of Mathematics, 155S 1400E, University of Utah, Salt Lake City, UT 84112, USA. Tel: +1 801 581 6822/6851; fax: +1 801 581 4181.}

\author[gd]{Grzegorz Dzier\.z{}anowski}
\ead{gd@il.pw.edu.pl}
\address[gd]{Faculty of Civil Engineering, Warsaw University of Technology, Warsaw, Poland.}

\begin{abstract}
The paper establishes exact lower bound on the effective elastic energy of two-dimensional, three-material composite subjected to the homogeneous, anisotropic stress. It is assumed that the materials are mixed with given volume fractions and that one of the phases is degenerated to void, i.e.\ the effective composite is porous. Explicit formula for the energy bound is obtained using the translation method enhanced with additional inequality expressing certain property of stresses. Sufficient optimality conditions of the energy bound are used to set the requirements which have to be met by the stress fields in each phase of optimal effective material regardless of the complexity of its microstructural geometry. We show that these requirements are fulfilled in a special class of microgeometries, so-called laminates of a rank. Their optimality is elaborated in detail for structures with significant amount of void, also referred to as \emph{high-porosity structures}. It is shown that geometrical parameters of optimal multi-rank, high-porosity laminates are different in various ranges of volume fractions and anisotropy level of external stress. Non-laminate, three-phase microstructures introduced by other authors and their optimality in high-porosity regions is also discussed by means of the sufficient conditions technique. Conjectures regarding low-porosity regions are presented, but full treatment of this issue is postponed to a separate publication. The corresponding ``G-closure problem'' of a three-phase isotropic composite is also addressed and exact bounds on effective isotropic properties are explicitly determined in these regions where the stress energy bound is optimal.
\end{abstract}

\begin{keyword}
Multimaterial composites \sep Minimal stress energy \sep Bounds for effective properties \sep Optimal microstructures \sep High-rank laminates 
\end{keyword}

\end{frontmatter}

\section{Introduction}
\label{sec1-intro}
\input{3mat-acgd-revised-sec1}

\section{Problem setting}
\label{sec2-setting}
\input{3mat-acgd-revised-sec2}

\section{Lower bound on the stress energy: Sufficient optimality condition}
\label{sec3-bound}
\input{3mat-acgd-revised-sec3}

\section{Optimal microstructures in high-porosity regions}
\label{sec4-laminates}
\input{3mat-acgd-revised-sec4}

\section{Remarks on low-porosity regions}
\label{sec5-remarks}
\input{3mat-acgd-revised-sec5}

\section*{Acknowledgements}
This work has been partially supported by the European Union in the framework of European Social Fund through the Warsaw University of Technology Development Programme (Grzegorz Dzier\.{z}anowski).

\bibliographystyle{elsarticle-harv}

\end{document}

%% file: 3mat-acgd-revised-sec1.tex
\subparagraph*{Significance of the problem}
Optimization of composite microstructures is important today because technological capabilities allow for manufacturing a huge variety of microscopic designs for roughly the same price, and one wants to know what ``the best'' microstructure is. There is no boundary between \emph{optimal structural design} in classical engineering sense and \emph{optimal composite material} as the latter is also a structure at microlevel. Optimal large-scale structures are made from optimal microstructures (composites) and the main difference between them is that the composite problem is solved for a periodic domain and periodic boundary conditions, which permits for an explicit solution. Besides the optimal structures, one wants to know the range of improvement of effective composite properties by varying the microstructure. The related quasiconvex envelope problem, see for example \citep{Che00, Dac08} opens ways to construction of metamaterials, i.e.\ structures with unusual responses.

So far, the vast majority of available results deals with two-material composites. Meanwhile, numerous applications call for optimal design of multimaterial composites, or even porous composites from two elastic materials and void. Especially worth noting are applications that utilize multi-physics, i.e. elastic and electromagnetic properties and those that deal with structures best adapted to variable environment such as natural morphologies perfected by evolution.

\begin{figure}[H]
\centering
\includegraphics[width=90mm]{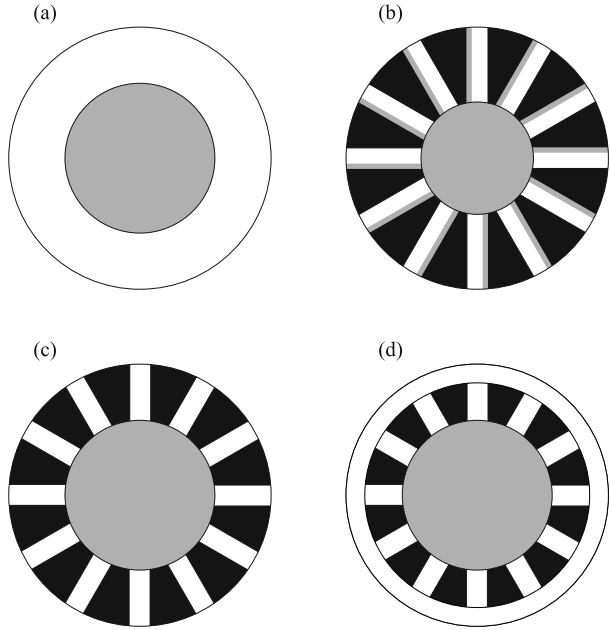}
\caption{Elements of a Hashin-Shtrikman type assemblage from two isotropic materials and void (white and grey colors represent stronger and weaker phases repectively, black corresponds to void). Comparison of geometries of optimal microstructures with maximal effective bulk modulus: (a) two-material composite; (b)-(d) three-material microstructures with small, intermediate and large volume of stronger material.\label{fig1}}
\end{figure}

Optimal microstructures of two-phase and multiphase composites are drastically different. In contrast with the steady and intuitively expected topology of two-material optimal mixture (a strong material always surrounds weak inclusions), optimal multimaterial structures show the large variety of patterns and the optimal topology depends on volume fractions. Optimal multiphase structure may contain an enveloping layer but it also has ``hubs'' of a material with intermediate stiffness connected by ``pathways'' (laminate of the best and worst materials) and other configurations that reveal a geometrical essence of optimality, see \citep{Che09,Che11,Che12} and Figure \ref{fig1}. Geometries of multimaterial optimal structures are not unique, pieces of the same material may occur in different places of an optimal structure and they may correspond to different fields inside them. Clearly, the method for finding optimal multiphase geometries differ from those for optimal two-material structures. 

In this paper we follow \citep{Che11} in developing the technique for finding the bounds and optimal structures and we apply it to elastic composites. The results constitute a next step from the popular ``topology optimization'', see \citep{Ben03}, that is a problem of optimal layout of a material and void. Namely, we describe optimal distributions of two materials and void, or optimal two-material porous composites made from a strong and expensive material, a weak and cheap one, and void. Such problem is useful for many multi-physics design applications when additional properties besides elasticity are important. The presence of one of the materials can be independently required by whatever reasons.

\subparagraph*{Background of the research}
Formally, the problem of optimal structures can be formulated as a question of minimizers of a variational problem with nonquasiconvex multiwell Lagrangians; the wells represent components' energies plus their costs and the minimizers (Young measures) are stress fields in the materials of an optimal composite. The challenging open problem is to build the quasiconvex envelope for Lagrangian with three or more wells. The problem is addressed by (i) finding exact bound (the lower bound for the quasiconvex envelope) and (ii) approximating these bounds by special class of minimizers. By building the lower bound, we also obtain sufficient conditions on optimal fields in materials that hint on the search for geometric patterns determining optimal structures, see \citep{Mil02,Alb07,Che09}.

The existing techniques for the bound such as Hashin-Shtrikman method, see \citep{Has63}; translation method, see \citep{Lur82,Lur86,Tar85,Mil02} or analytic method of Bergman-Milton, see \citep{Ber78,Mil81a,Mil02}, produced a number of results for two-material mixtures in the last 25 years, see e.g. the books \citep{Lur93,Che00,All02,Mil02,Dac08} for examples. These techniques, however, do not provide all solutions for multiwell problems.

In the last three decades, the multimaterial optimal composites have been studied by Milton \citep{Mil81b}, Lurie \& Cherkaev \citep{Lur85}, Milton \& Kohn \citep{Mil88} among others. In 1995, Nesi published a paper \citep{Nes95} about bounds for multimaterial mixtures that are better than Hashin-Shtrikman ones. Several new types of three-phase structures with bulk modulus equal or close to the Hashin-Shtrikman bound were suggested by Gibiansky \& Sigmund, see \citep{Gib00,  Sig00}.

In the last years (2009-2012), a new technique  for finding optimal bounds for multimaterial mixtures was suggested and tested on a couple of examples \citep{Che09, Che11, Che12}. The essence of the new technique is coupling the translation method with the Alessandrini-Nesi inequality, see \citep{Ale01}, that order and restrain values of the fields in any optimal composite. Roughly speaking, in the case tackled in this paper (elastic 2D microstructures of maximal stiffness for a mixture of two materials and void), the Alessandrini-Nesi inequality states that the sign of a stress field is constant in the whole microstructure. The technique was used to find the bounds on the effective properties of isotropic 2D multimaterial composites, see \citep{Che09,Che12} and anisotropic conducting composites made from two materials and void, see \citep{Che11}. 

The lower bound on the effective energy is a multifaceted surface, its analytic expression is different in different regions of volume fractions of mixed materials and anisotropy level of average stress. In this paper, optimal energy bound and locally optimal stress fields are analyzed and described for high-porosity composites, i.e.\ mixtures containing a significant amount of void. Low-porosity case is also addressed, but the detailed description is postponed to a separate publication.  In one region the optimality is conjectured. Our guess is that another, yet unaccounted, inequality becomes active and improves the bound in this region.

%% file: 3mat-acgd-revised-sec2.tex
\subsection{Notation}
\label{sec21-notation}

\subparagraph*{Reference to periodic homogenization}
Consider a domain $\Omega\subset\mR^2$ filled with two linearly elastic materials and a void. Non-homogeneous distribution of phases in the domain is determined by its division into three disjoint subsets $\Omega_i$, $i=1,2,3$. Suppose that a boundary value problem (BVP) of linearized elasticity is posed in $\Omega$. If non-homogeneity of a material layout is given by a fine partition of the domain then it is convenient to make use of the homogenization theory of periodic media in determining the simplified, effective Hooke's law in $\Omega$ prior to solving the BVP.

In this paper we solve an inverse homogenization problem: we find a structure of a multicomponent composite that stores minimal stress energy in a given homogeneous stress field. We assume that the properties of constituent materials and their volume fractions are given.

\subparagraph*{Definitions}
Due to the local character of homogenization, in the sequel we consider arbitrary $x\in\Omega$ which is sufficiently distant from the boundary $\partial\Omega$. Let $Y=[0,1]^2$ denote a corresponding unit cell periodically extended to $\mR^2$. Assume that $Y$ is divided into three disjont subcells $Y_i$, $i=1,2,3$, whose areas $m_i$ are fixed. Write
\be
\label{e21-1}
Y = \bigcup\limits_{i=1,2,3} Y_i, \quad |Y_i| = m_i, \quad \sum_{i=1}^3 m_i = 1
\ee
and set $(e_1, e_2)$ for a Cartesian basis in $Y$. Let $\mE^2_s$ stand for a space of plane, second-order symmetric tensors, and $\mE^4_s$ for the space of plane Hooke's tensors. Next, choose
\be
\label{e21-1a}
\ba{ll}
E_1 = \dfrac{1}{\sqrt{2}}(e_1\otimes e_1 + e_2\otimes e_2), &E_2 = \dfrac{1}{\sqrt{2}}(e_1\otimes e_1 - e_2\otimes e_2),\\\noalign{\smallskip}
E_3 = \dfrac{1}{\sqrt{2}}(e_1\otimes e_2 + e_2\otimes e_1)
\ea
\ee
for the basis in $\mE^2_s$.

Suppose that $Y_1$ and $Y_2$ are filled with elastic isotropic materials whose constitutive properties are given by $K_i=1/\kappa_i$, $L_i=1/\mu_i$, $i=1,2$, where $\kappa_i$ and $\mu_i$ stand for bulk and shear moduli of $i$-th phase. Let $K_3=L_3=+\infty$ which means that the third phase corresponds to void.  Introduce a set $A=\{A_1, A_2, A_\mrm{void}\}$ where
\be
\label{e21-1b}
A_i = \dfrac{K_i}{2}\,E_1\otimes E_1 + \dfrac{L_i}{2}\,(E_2\otimes E_2 + E_3\otimes E_3)
\ee
represents Hooke's compliance tensor of $i$-th non-degenerate isotropic phase. In the sequel we assume that the materials are well-ordered, i.e.\ $K_1 < K_2 < K_3=+\infty$ and $L_1 < L_2 < L_3=+\infty$.

Set
\[
\tau_0 = \eta\,e_1\otimes e_1 + \varrho\,e_2\otimes e_2
\]
for the average stress tensor in $Y$. Components $\eta$ and $\varrho$ denote principal values of $\tau_0$ and $(e_1, e_2)$ stands for its principal basis. We normalize $\tau_0$, assuming without loss of generality $\eta=1$, $|\varrho|\leq 1$. It follows that
\[
\tau_0 = S_0\,E_1 + D_0\,E_2,\quad S_0 = \dfrac{1+\varrho}{\sqrt{2}},\quad D_0 = \dfrac{1-\varrho}{\sqrt{2}}
\]
and $S_0$, $D_0$ represent spherical and deviatoric components of $\tau_0$.

Stress fields satisfy equlibrium condition $\mbox{div}\,\tau = 0$. We define a set of statically admissible stress fields in $Y$
\[
\Sigma = \Bigg\{\tau\colon\tau\in L_\#^2(Y,\mE^2_s),\ \mbox{div}\,\tau = 0 \mbox{ in } Y,\ \int_Y \tau(y) dy = \tau_0 \Bigg\}
\]
where $L_\#^2(Y,\mE^2_s)$ stands for the space of $L^2$-functions with values in $\mE^2_s$ and $Y$-periodic in $\Omega$. 

Due to $Y$-periodicity, $\tau\in\Sigma$ is endowed with two properties:
\begin{itemize}
\item[--] function $\det\tau(y)$, $y\in Y$, is quasiaffine hence
\be
\label{e21-5}
\int_Y \det\tau(y) dy = \det\tau_0 = \varrho ;
\ee
\item[--] function $\det\tau(y)$, $y\in Y$, is locally univalent with $\det\tau_0$, that is 
\be
\label{e21-6}
\det\tau(y)\geq 0\ \mbox{a.e. in $Y$ if}\ \det\tau_0\geq 0
\ee
and the latter remains valid if $``\geq"$ is replaced by $``\leq"$ ,
\end{itemize}
see \citep{Ale01}. The above-mentioned properties do not result in any restrictions on $\tau\in\Sigma$, they simply unveil certain characteristics of the stress fields related to assumed $Y$-periodicity. Nevertheless, \eqref{e21-5} and \eqref{e21-6} are of great significance in bounding the stress energy which is the central part of the study. 

Symmetric second order tensor $\tau$ is uniquely represented in \eqref{e21-1a} by one spherical and two deviatoric components, respectively given by $s$ and $d_1$, $d_2$, such that
\[
s = \dfrac{\tau_{11}+\tau_{22}}{\sqrt{2}},\quad d_1 =  \dfrac{\tau_{11}-\tau_{22}}{\sqrt{2}},\quad d_2 = \sqrt{2}\,\tau_{12}
\]
hence $\tau(y) = s(y)\,E_1 + d_1(y)\,E_2 + d_2(y)\,E_3$. Decomposing the determinant function of a stress field according to 
\[
2\,\det\tau = s^2 - \left(d_1\2 +d_2\2\right)
\]
and considering $\varrho\in [-1, 1]$, allows for rewriting \eqref{e21-6} in the form
\be
\label{e21-8}
\ba{l}
s^2(y) \geq d_1\2(y) + d_2\2(y)\ \mbox{a.e. in $Y$ if}\ \varrho\in [0,\ 1],\\\noalign{\smallskip}
s^2(y) \leq d_1\2(y) + d_2\2(y)\ \mbox{a.e. in $Y$ if}\ \varrho\in [-1, 0].
\ea
\ee

For further considerations, let us rephrase the requirements imposed on $\tau\in\Sigma$. First, define a set
\[
\Sigma_{\mrm{uni}} = \Big\{\tau\colon\tau\in L_\#^2(Y,\mE^2_s)\ \mbox{with univalence property as in \eqref{e21-8}},\Big\} .
\]
Next, write the restriction on the average stress ($\int_Y \tau = \tau_0$) in a form
\[
\ba{lll}
\Sigma_{\mrm{av}} = &\Big\{\ S_i, D_{ij},\ i,j=1,2\colon & m_1 S_1 + m_2 S_2 = S_0,\\\noalign{\smallskip}
&\ \ \ \ m_1 D_{11} + m_2 D_{12} = D_0, & m_1 D_{12} + m_2 D_{22} = 0,\\\noalign{\medskip}
&\ \ \ \ S_i\2\geq D_{i1}\2 + D_{i2}\2, &\mbox{if $\varrho\in[0,\ 1]$},\\\noalign{\smallskip}
&\ \ \ \ S_i\2\leq D_{i1}\2 + D_{i2}\2, &\mbox{if $\varrho\in[-1,0]$}\ \ \Big\}
\ea
\]
where 
\be
\label{e21-11}
\ba{lll}
S_i = \dfrac{1}{m_i}\disp\int_{Y_i} s(y)\,dy, &D_{ij} = \dfrac{1}{m_i}\disp\int_{Y_i} d_j(y)\,dy, &i,j=1,2,
\ea
\ee
denote average spherical and deviatoric stresses in non-degenerate phases. 

It follows that $\Sigma\subseteq\Sigma_{\mrm{rel}}$ where
\be
\label{e21-13}
\Sigma_{\mrm{rel}} = \Big\{\tau\colon\tau\in\Sigma_{\mrm{uni}}\ \mbox{and such that }S_i, D_{ij}\in\Sigma_{\mrm{av}},\ i,j=1,2\Big\}
\ee
stands for a set of relaxed stress fields, i.e.\ fields with neglected equlibrium condition $\mrm{div}\tau = 0$ in $Y$.

\subsection{Composite materials of minimal stress energy}
\label{sec22-minimum}

\subparagraph*{Energy bound and extremal effective material properties}
The (quadrupled) stress energy density in $Y_i$, $i=1,2$, is calculated according to
\be
\label{e22-1}
U_i(\tau) = 4\,\big[\tau:(A_i\,\tau)\big] = K_i\,s^2 + L_i\,\left(d_1\2 + d_2\2\right)
\ee
and we set $U_3(\tau) = 0$ due to assumed $\tau=0$ in void. The contraction $\tau:(A_i\,\tau)$ is realized by a standard operation $[\tau]^T\,(A_i)\,[\tau]$ in the basis \eqref{e21-1a}. Here $[\tau]$ and $(A_i)$ stand for a vector and matrix representations of respective quantities and $[\tau]^T$ denotes a transpose of $[\tau]$. Effective energy is thus calculated according to
\be
\label{e22-2}
U_0(\varrho) = \inf \bigg\{\int_{Y_1} U_1(\tau)\,dy + \int_{Y_2} U_2(\tau)\,dy\ \bigg|\ \tau\in\Sigma \bigg\}
\ee
and $U_0(\varrho)$ is bounded from below by
\[
U_\ast(\varrho) = \inf\Big\{ U_0(\varrho)\ \Big|\ Y_i\mbox{ as in \eqref{e21-1}}\Big\} .
\]

Bounding the stress energy allows for restricting the values of effective constitutive properties. Indeed, by introducing $K_\ast$, $L_\ast$ and $A_\ast$ linked similarly to \eqref{e21-1b} one may claim $U_\ast(\varrho)$ in the form
\be
\label{e22-4}
U_\ast(\varrho) = 4\,\big[\tau_0:(A_\ast\,\tau_0)\big] = K_\ast\,S_0\2+L_\ast\,D_0\2 = \dfrac{1}{2}\bigg(K_\ast\,(1+\varrho)^2 + L_\ast\,(1-\varrho)^2\bigg) .
\ee
With this notation, $K_\ast$ and $L_\ast$ represent coupled bounds on effective moduli of a composite for fixed $\varrho$. They may be understood as constitutive properties of a homogenized medium adjusted to the external stress $\tau_0 = S_0\,E_1 + D_0\,E_2$ in a sense of storing the minimal amount of energy in two directions $E_1$, $E_2$ simultaneously.

Note that the requirement of isotropy imposed on the effective me\-dium is redundant. Indeed, the component of $A_\ast$ related to the direction $E_3\otimes E_3$ may be arbitrary as $\tau_0:E_3 = 0$. Non-isotropic microstructres may thus be optimal, i.e.\ such that the amount of stress energy stored in them equals $U_\ast(\varrho)$. Details on this topic are presented in Sec.\ \ref{sec4-laminates}.

Let us find formulae for $K_\ast$ and $L_\ast$. To this end, note that by varying $\varrho\in [-1,1]$ on the r.h.s.\ of \eqref{e22-4} we obtain a family of functions that are quadratic in $\varrho$ and $U_\ast(\varrho)$ represents an envelope of this family. Solving the system
\[
\ba{l}
\phantom{\dfrac{d}{d\varrho}\Big[}U_\ast(\varrho) - \dfrac{1}{2}\bigg(K_\ast\,(1+\varrho)^2 + L_\ast\,(1-\varrho)^2\bigg) = 0,\\\noalign{\smallskip}
\dfrac{d}{d\varrho}\bigg[U_\ast(\varrho) - \dfrac{1}{2}\bigg(K_\ast\,(1+\varrho)^2 + L_\ast\,(1-\varrho)^2\bigg)\bigg] = 0,
\ea
\]
allows for determining the coefficients of $U_\ast(\varrho)$. They read
\be
\label{e22-6}
\ba{l}
K_\ast(\varrho) = \dfrac{U_\ast(\varrho)}{1+\varrho}+\dfrac{1-\varrho}{2(1+\varrho)}\dfrac{dU_\ast(\varrho)}{d\varrho},\\\noalign{\smallskip}
L_\ast(\varrho) = \dfrac{U_\ast(\varrho)}{1-\varrho}-\dfrac{1+\varrho}{2(1-\varrho)}\dfrac{dU_\ast(\varrho)}{d\varrho}  .
\ea
\ee

Functions in \eqref{e22-6} are extremal if their values belong to $\partial G_mA$, i.e.\ the boundary of $G$-\emph{closure} of set $A$. Recall that $G_mA$ contains all effective Hooke's tensors obtained by homogenization of components belonging to $A$, taken with arbitrary microstructure and fixed volume fractions $m_i$, see e.g.\ \citep{Che00} for further reference.

\subparagraph*{Calculating energy bound by the translation method}
In what follows we briefly describe a procedure of determining $U_\ast(\varrho)$. For this we make use of \emph{the translation method} which proved to be an efficient tool in solving problems regarding energy and effective property bounds posed in various settings, see \citep{Che00,Mil02}. The method starts from introducing \emph{a translation parameter} $\alpha\in\mT\subset\mR$ and rephrasing \eqref{e22-1} in the form
\[
U_i(\tau) = F_i(\tau,\alpha) - 2\,\alpha\det\tau,\qquad i=1,2,
\]
where
\[
F_i(\tau,\alpha) = (K_i+\alpha)\,s^2 + (L_i-\alpha)\left(d_1\2 +d_2\2\right) .
\]

With \eqref{e21-5} taken into consideration we calculate
\[
\int_{Y_1} U_1(\tau)\,dy + \int_{Y_2} U_2(\tau)\,dy = \int_{Y_1} F_1(\tau,\alpha)\,dy + \int_{Y_2} F_2(\tau,\alpha)\,dy - 2\,\varrho\,\alpha.
\]

Next, we neglect the differential constraint $\mrm{div}\tau = 0$ on the stress field in $Y$. This reduces the problem to an algebraic one and allows for taking the infimum in \eqref{e22-2} on the enlarged set $\Sigma_{\mrm{rel}}$. Optimal stress field $\tau\in\Sigma_{\mrm{rel}}$ can be now determined independently in each phase which also follows from dropping $\mrm{div}\,\tau = 0$ in $Y$. The search is reduced to non-degenarate phases only as $\tau=0$ in void. Consequently, one obtains
\be
\label{e22-9}
\ba{l}
U_0(\varrho)\geq \Phi(\varrho,\alpha) - 2\,\varrho\,\alpha,\\\noalign{\smallskip}
\Phi(\varrho,\alpha) = \inf\bigg\{\disp\int_{Y_1} F_1(\tau,\alpha)\,dy + \disp\int_{Y_2} F_2(\tau,\alpha)\,dy\ \bigg|\ \tau\in\Sigma_{\mrm{rel}}\bigg\} .
\ea
\ee

By \eqref{e21-13} it is possible to split the latter task into two steps. First, we define the energy function $\Phi_i$ in the domain $Y_i$
\be
\label{e22-10}
\Phi_i(S_i, D_{i1}, D_{i2},\alpha) = \inf \bigg\{\int_{Y_i} F_i(\tau,\alpha)\,dy\ \bigg|\ \tau\in \Sigma_{\mrm{uni}} \bigg\},\quad i=1,2,
\ee
finding the best distribution of $\tau$ within $Y_i$. Then we continue with
\be
\label{e22-11}
\Phi(\varrho,\alpha) = \min\bigg\{\Phi_1 + \Phi_2\ \bigg|\ S_i, D_{ij}\in\Sigma_{\mrm{av}}\bigg\}
\ee
that describes the distribution of $\tau$ in the whole $Y$. Finally, we choose translation parameter $\alpha$, obtaining the best lower bound on the stress energy
\be
\label{e22-12}
U_\ast(\varrho) \geq U_{\mrm{tr}}(\varrho) = \max\Big\{\Phi(\varrho,\alpha) - 2\,\varrho\,\alpha\ \Big|\ \alpha\in\mT\Big\} .
\ee

The equality $U_\ast(\varrho) = U_{\mrm{tr}}(\varrho)$ holds if the minimizer $\tau\in\Sigma_{\mrm{rel}}$ is statically admissible, i.e.\ $\mrm{div}\,\tau = 0$ in $Y$, $\tau\in\Sigma$. If this is the case then the bound $U_{\mrm{tr}}(\varrho)$ is \emph{optimal}, or \emph{exact}, as it corresponds to the boundary of $G_mA$ and it may be substituted in \eqref{e22-6} for calculating extremal coupled effective properties of a three-phase composite. Explicit calculation of $U_{\mrm{tr}}(\varrho)$ is a subject of Sec.\ \ref{sec3-bound}, and proving its optimality is postponed until Sec.\ \ref{sec4-laminates}.

%% file: 3mat-acgd-revised-sec3.tex
\begin{figure}[H]
\centering
\includegraphics[width=\textwidth]{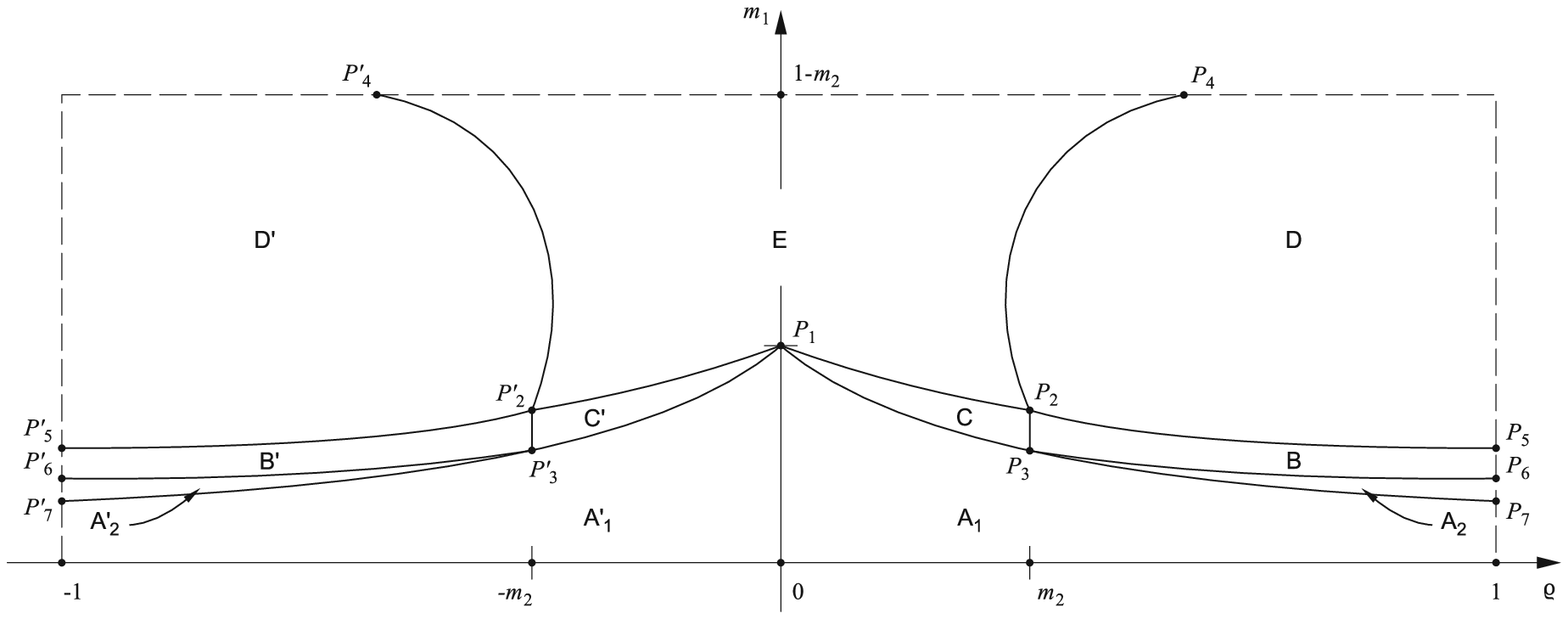}
\caption{Regions of optimality related to a cross-section of a polyhedron $\Pi$ by a plane $m_2=0.35$. Constitutive properties of nondegenerate materials are fixed to $K_1=1$, $L_1=2$, $K_2=3$, $L_2=4$.\label{fig2}}
\end{figure}

We proceed by explicit calculation of $\tau\in\Sigma_{\mrm{rel}}$ in two steps defined by \eqref{e22-10} and \eqref{e22-11}. This in turn allows for determining $U_{\mrm{tr}}(\varrho)$ by proper adjustment of the translation parameter $\alpha$ in \eqref{e22-12}. Consequently, bounds on effective constitutive properties $K_\ast(\varrho)$ and $L_\ast(\varrho)$ are obtained through \eqref{e22-6}. These bounds are exact if the energy bound is exact, i.e.\ when $U_\ast(\varrho) = U_{\mrm{tr}}(\varrho)$ holds. Discussion of the latter is provided in Sections \ref{sec4-laminates} and \ref{sec5-remarks}. Below we establish \emph{the sufficient optimality condition} in terms of stress fields related to $U_{\mrm{tr}}(\varrho)$. With $K_i$, $L_i$, $i=1,2$ given, the sought condition turns out to be dependent on mutual relations among $m_1$, $m_2$ and $\varrho$. It results in the division of a polyhedron $\Pi=\{(\varrho, m_1, m_2): \varrho\in[-1,1], m_1\in[0,1-m_2], m_2\in[0,1]\}$ into several regions. Table \ref{tab1} provides a brief guide to the sequence and results of calculations and Fig.\ \ref{fig2} shows an exemplary cross-section of $\Pi$ by a plane $m_2=\mrm{const.}$

\begin{table}[H]
\caption{A guide to the results of calculations of the exact lower bound on the stress energy, optimal effective isotropic properties and optimal fields in materials. \label{tab1}}
\vspace*{3pt} 
\begin{tabular*}{\textwidth}{@{\extracolsep{\stretch{1}}}lllll}
\hline\noalign{\smallskip}
range of $\varrho$ &region$\,^\ast$ &$U_{\mrm{tr}}(\varrho)$ &$K_\ast(\varrho)$,  $L_\ast(\varrho)$ &optimal fields\\
\noalign{\smallskip}\hline\noalign{\smallskip}
$\varrho\in[-1,0]$ &$\cA'$ &\eqref{e324-4} &\eqref{e33-2} &\eqref{e31-10}, \eqref{e324-5} -- phase 1\\
                                                                                                                           &&&&\eqref{e31-11}, \eqref{e324-5} -- phase 2\\
                                         &$\cB'$ &\eqref{e323-6} &\eqref{e33-4} & \eqref{e31-10}, \eqref{e323-8} -- phase 1\\
                                                                                                                           &&&& \eqref{e31-3}, \eqref{e323-8} -- phase 2 \\
                                         &$\cC'$ &\eqref{e323-13} &\eqref{e33-5} & \eqref{e31-10}, \eqref{e323-15} -- phase 1\\
                                                                                                                           &&&& \eqref{e31-3}, \eqref{e323-15} -- phase 2\\
                                         &$\cD'$ &\eqref{e325-12} &\eqref{e33-7} &\eqref{e31-11}, \eqref{e325-13} -- phase 1\\
                                                                                                                           &&&& \eqref{e31-3}, \eqref{e325-13} -- phase 2\\
\noalign{\smallskip}\hline\noalign{\smallskip}
$\varrho\in[0,\ 1]$&$\cD$  &\eqref{e325-10} &\eqref{e33-6} &\eqref{e31-7}, \eqref{e325-11} -- phase 1\\
                                                                                                                           &&&& \eqref{e31-3}, \eqref{e325-11} -- phase 2\\
                                         &$\cC$   &\eqref{e321-13} &\eqref{e33-5} &\eqref{e31-6}, \eqref{e321-15} -- phase 1\\
                                                                                                                            &&&& \eqref{e31-3}, \eqref{e321-15} -- phase 2\\
                                         &$\cB$   &\eqref{e321-6} &\eqref{e33-3} &\eqref{e31-6}, \eqref{e321-8} -- phase 1\\
                                                                                                                            &&&& \eqref{e31-3}, \eqref{e321-8} -- phase 2\\
                                         &$\cA$   &\eqref{e322-4} &\eqref{e33-1} &\eqref{e31-6},  \eqref{e322-5} -- phase 1\\
                                                                                                                             &&&&\eqref{e31-7},  \eqref{e322-5} -- phase 2\\
\noalign{\smallskip}\hline\noalign{\smallskip}
                &$\cE$   &-----   &----- &-----\\
\noalign{\smallskip}\hline
\end{tabular*}

\vspace{3 pt}

$^\ast$ see Fig.\ \ref{fig2}
\end{table}

\subsection{Requirements for optimal stress fields in phases}
\label{sec31-fields}

In \eqref{e22-10} we wish to obtain $\Phi_i(S_i, D_{i1}, D_{i2},\alpha)>-\infty$ for $i=1,2$, as such property is crucial in subsequent derivation of a nontrivial energy bound $U_{\mrm{tr}}(\varrho)$. To this end, we first discuss the conditions under which the integrand $F_i(\tau,\alpha)$ is bounded from below by a convex function. Next, we set the requirements for optimal relaxed stress fields $\tau\in\Sigma_{\mrm{uni}}$ by making use of the Jensen inequality, see \citep[Sec.\ 1.2]{Che00}. Applied to our case, it states that if $F_i(\tau,\alpha)$ is convex in $\tau$ then its integral over $Y_i$ takes a minimum value on a constant stress field being the average of $\tau$ over $Y_i$. Thus, with $\tau$ decomposed into spherical and deviatoric parts, we expect the minimizers to be expressed in terms of averages $S_i, D_{i1}, D_{i2}$, see \eqref{e21-11}. 

Assuming that $S_i, D_{ij}$, $i,j=1,2$, are prescribed and $K_i\neq L_i$, we consider the following cases:
\begin{enumerate}
\item[(I)] If $\alpha\in(-K_i, L_i)$ then all terms in $F_i(\tau,\alpha)$ are convex.

For $\varrho\in [-1,1]$, from the Jen\-sen inequality it follows that
\[
\Phi_i(S_i, D_{i1}, D_{i2},\alpha) = m_i(K_i+\alpha)\,S_i\2 +m_i(L_i-\alpha)\left(D_{i1}\2 +D_{i2}\2\right).
\]
Components of optimal relaxed field $\tau\in\Sigma_{\mrm{uni}}$ are constant in $Y_i$ :
\be
\label{e31-3}
s(y) = S_i,\quad d_j(y) = D_{ij}\quad\mbox{ a.e. in $Y_i$}.
\ee

\item[(IIa)] If $\alpha > L_i$ then the $s$-term in $F_i(\tau,\alpha)$ is convex and $d$-terms are concave.

For $\varrho\in [0,1]$ we have $s^2\geq d_1\2 +d_2\2$ a.e.\ in $Y_i$ hence $F_i(\tau,\alpha)$ is bounded from below by a convex function. Indeed, 
\[
(K_i+\alpha)\,s^2+(L_i-\alpha)\left(d_1\2 +d_2\2\right)\geq (K_i+L_i)\,s^2
\]
and the $``\geq"$ relation above is replaced by $``="$ when $s^2 = d_1\2 +d_2\2$ a.e.\ in $Y_i$. Consequently,
\be
\label{e31-5}
\Phi_i(S_i, D_{i1}, D_{i2},\alpha) = m_i(K_i+L_i)\,S_i\2.
\ee
Components of optimal $\tau\in\Sigma_{\mrm{uni}}$ are given by  
\be
\label{e31-6}
\ba{l}
s(y) = S_i\ \mbox{(constant field)},\\\noalign{\smallskip}
d_1(y) =S_i\cos\theta(y),\ \ d_2(y)=S_i\sin\theta(y),\ \ \mbox{a.e.\ in $Y_i$}
\ea
\ee
with $\theta(y)$ arbitrary in $Y_i$ up to the restrictions given in \eqref{e21-8}.

\item[(IIb)] If $\alpha = L_i$ then the $s$-term in $F_i(\tau,\alpha)$ is convex and $d$-terms vanish.

For $\varrho\in [0,1]$ we obtain $\Phi_i(S_i, D_{i1}, D_{i2},\alpha)$ in a form identical to \eqref{e31-5}. Formulae determining the components of optimal relaxed field read
\be
\label{e31-7}
\ba{l}
s(y) = S_i\ \mbox{(constant field)},\\\noalign{\smallskip}
d_j(y)\ \mbox{ arbitrary up to $s^2\geq d_1\2+d_2\2$ a.e. in $Y_i$}.
\ea
\ee

\item[(IIIa)] If $\alpha < -K_i$ then the $s$-term in $F_i(\tau,\alpha)$ is concave and $d$-terms are convex. 

For $\varrho\in [-1,0]$ we have $s^2\leq d_1\2 +d_2\2$ a.e.\ in $Y_i$ hence
\be
\label{e31-8}
(K_i+\alpha)\,s^2+(L_i-\alpha)\left(d_1\2 +d_2\2\right)\geq (K_i+L_i)\left(d_1\2 +d_2\2\right)
\ee
and the $``\geq"$ relation above is replaced by $``="$ when $s^2 = d_1\2 +d_2\2$ a.e.\ in $Y_i$. Consequently,
\be
\label{e31-9}
\Phi_i(S_i, D_{i1}, D_{i2},\alpha) = m_i(K_i+L_i)\,\left(D_{i1}\2 +D_{i2}\2\right).
\ee
Components of optimal relaxed stress field are constant in $Y_i$ :
\be
\label{e31-10}
s(y)=\pm\sqrt{D_{i1}\2 +D_{i2}\2},\quad d_j(y)=D_{ij}\ \mbox{a.e.\ in $Y_i$}.
\ee

\item[(IIIb)] If $\alpha = -K_i$ then the $s$-term in $F_i(\tau,\alpha)$ vanishes and $d$-terms are convex. 

For $\varrho\in [-1, 0]$ we obtain $\Phi_i(S_i, D_{i1}, D_{i2},\alpha)$ in the form identical to \eqref{e31-9} with optimal relaxed fields
\be
\label{e31-11}
\ba{l}
s(y)\mbox{ arbitrary up to $s^2\leq d_1\2 +d_2\2$, a.e. in $Y_i$},\\\noalign{\smallskip}
d_j(y) = D_{ij}\ \mbox{(constant field)}.
\ea
\ee
\end{enumerate}
Other relations between $\varrho$ and $\alpha$ are not discussed here as they are irrelevant in further study.


\subsection{Calculation of the energy bound: Regions of optimality}
\label{sec32-regions}

Having $\Phi_i = \Phi_i(S_i, D_{i1}, D_{i2},\alpha)$, $i=1,2$, explicitly calculated, we now turn to the problem of determining $\Phi(\varrho, \alpha)$ and $U_{\mrm{tr}}(\varrho)$ through \eqref{e22-11} and \eqref{e22-12}. Substituting thus obtained optimal $S_{i}$, $D_{i1}$, $D_{i2}$ in the formulae for $s(y)$ and $d_1(y)$, $d_2(y)$ leads to the explicit form of the requirements for optimal relaxed stress fields derived in previous Section.

\subsubsection{Case of $\varrho\in [0,1]$ and $\alpha\in (L_1, L_2)$: Regions $\mathsf{B}$ and $\mathsf{C}$}
\label{sec321-regionBC}

According to the discussion in Sec. \ref{sec31-fields}, items (I) and (IIa), set
\[
\ba{l}
\Phi_1 = m_1(K_1+L_1)\,S_1\2,\\\noalign{\smallskip}
\Phi_2 = m_2(K_2+\alpha)\,S_2\2 +m_2(L_2-\alpha)\left(D_{21}\2 +D_{22}\2\right) .
\ea
\]
It follows that
\be
\label{e321-2}
\ba{ll}
\Phi(\varrho,\alpha) = &\min\big\{ \Phi_1 + \Phi_2 \big\}\\\noalign{\smallskip}
&\mbox{subject to:}\ m_1S_1+m_2S_2 = S_0,\\\noalign{\smallskip}
&\phantom{subject to:}\ m_1D_{11}+m_2D_{21} = D_0,\\\noalign{\smallskip}
&\phantom{subject to:}\ m_1D_{12}+m_2D_{22} = 0,\\\noalign{\smallskip}
&\phantom{subject to:}\ S_1\2\geq D_{11}\2 + D_{12}\2,\\\noalign{\smallskip}
&\phantom{subject to:}\ S_2\2\geq D_{21}\2 + D_{22}\2.
\ea
\ee

From the KKT optimality conditions we conclude that: (i) $D_{12}=D_{22}=0$, (ii) $\Phi_1+\Phi_2$ is minimized with respect to $D_{21}$ if $D_{11}$ is maximized. All constraints in \eqref{e321-2} are satisfied if we set
\[
D_{11} = \min\left\{\dfrac{1}{m_1}D_0,\ S_1\right\} .
\]
Consequently, further discussion splits into two subcases corresponding to regions of optimality $\cB$ and $\cC$ in Fig.\ \ref{fig2}. Forthcoming results are rather straightforward to obtain in both regions, hence we omit the details of calculations. 

\subparagraph*{Region $\cB$\ :} Assume that
\[
D_{11} = \dfrac{1}{m_1}D_0\leq S_1.
\]
By this, $D_{21}=0 $ and \eqref{e321-2} transforms to
\be
\label{e321-5}
\ba{ll}
\Phi(\varrho,\alpha) = &\min\Big\{m_1(K_1+L_1)\,S_1\2 + m_2(K_2+\alpha)\,S_2\2\Big\}\\\noalign{\smallskip}
&\mbox{subject to:}\quad m_1S_1+m_2S_2 = S_0,
\ea
\ee
hence one may replace the above with a one-dimensional unconstrained optimization problem where $S_1$ is treated as a variable. The necessary optimality condition $\partial (\Phi_1+\Phi_2)/\partial S_1 = 0$ allows for determining the function $S_1(\alpha)$ which is substituted back in \eqref{e321-5}, and formula for $S_2(\alpha)$ follows from the constraint. The lower bound on stress energy in region $\cB$ given by
\be
\label{e321-6}
\ba{ll}
U_{\mrm{tr}}(\varrho) &= \max\Big\{\Phi(\varrho,\alpha) - 2\,\varrho\,\alpha\ \Big|\ \alpha\in(L_1, L_2)\Big\} =\\\noalign{\smallskip}
&= \dfrac{(1+\varrho-2\sqrt{\varrho\,m_2})^2}{2\,m_1}(K_1+L_1) + 2\varrho K_2
\ea
\ee
results from solving $\partial[\Phi(\varrho,\alpha)-2\,\varrho\,\alpha]/\partial\alpha = 0$. In this way, two critical values of $\alpha$ are obtained. The one corresponding to maximum in \eqref{e321-6} reads
\[
\alpha = \dfrac{1}{2}\,\dfrac{\sqrt{\varrho\,m_2}(1+\varrho-2\sqrt{\varrho\,m_2})}{\varrho\,m_1}(K_1+L_1)-K_2 .
\]
Consequently, optimal average spherical and deviatoric stress components in region $\cB$ are given by 
\be
\label{e321-8}
\ba{llll}
\mbox{in phase $1$:}
&S_1 = \dfrac{1+\varrho-2\sqrt{\varrho\,m_2}}{\sqrt{2}\,m_1},
&D_{11} = \dfrac{1-\varrho}{\sqrt{2}\,m_1},
&D_{12} = 0,\\\noalign{\smallskip}
\mbox{in phase $2$:}
&S_2 = \sqrt{2}\,\dfrac{\sqrt{\varrho\,m_2}}{m_2} ,
&D_{21} = 0, 
&D_{22} = 0.
\ea
\ee 

Sufficient optimality condition of $U_\mrm{tr}$ in region $\cB$ expressed in terms of stress fields in phases 1 and 2 respectively follows from substituting $S_1, D_{11}, D_{12}$ in \eqref{e31-6} and $S_2, D_{21}, D_{22}$ in \eqref{e31-3}.

Region $\cB$ is represented by a curvilinear rectangle $P_2P_3P_5P_6$ in Fig.\ \ref{fig2}. Its boundaries are determined according to the following scheme
\[
\ba{lll}
\alpha < L_2 &\Rightarrow &m_1 > \psi_{\cA-\cB}(m_2,\varrho),\\\noalign{\smallskip}
\alpha > L_1 &\Rightarrow &m_1 < \psi_{\cB-\cD}(m_2,\varrho),\\\noalign{\smallskip}
m_1S_1\geq D_0 &\Rightarrow &\varrho\in [m_2,1]
\ea
\]
where
\be
\label{e321-10}
\ba{l}
\psi_{\cA-\cB}(m_2,\varrho) = \dfrac{\sqrt{\varrho\,m_2}(1+\varrho-2\sqrt{\varrho\,m_2})}{2\,\varrho}\,\dfrac{K_1+L_1}{K_2+L_2},\\\noalign{\smallskip}
\psi_{\cB-\cD}(m_2,\varrho) = \dfrac{\sqrt{\varrho\,m_2}(1+\varrho-2\sqrt{\varrho\,m_2})}{2\,\varrho}\,\dfrac{K_1+L_1}{K_2+L_1}.
\ea
\ee

\subparagraph*{Region $\cC$\ :} Conversely to the previous paragraph assume
\[
D_{11} = S_1\leq \dfrac{1}{m_1}D_0,
\]
which results in
\[
\ba{ll}
\Phi(\varrho,\alpha) = &\min\Big\{m_1(K_1+L_1)\,S_1\2 +\\
&\phantom{\min} + m_2\big[(K_2+\alpha)\,S_2\2 + (L_2-\alpha)D_{21}\2\big]\Big\}\\\noalign{\smallskip}
&\mbox{subject to:}\ m_1S_1+m_2S_2 = S_0,\\\noalign{\smallskip}
&\phantom{subject to:}\ m_1S_1+m_2D_{21} = D_0 .
\ea
\]
The algorithm of calculations is similar to the one presented for region $\cB$. It follows that the stress energy in region $\cC$ is bounded from below by
\be
\label{e321-13}
\ba{ll}
U_{\mrm{tr}}(\varrho) &= \max\Big\{\Phi(\varrho,\alpha) - 2\,\varrho\,\alpha\ \Big|\ \alpha\in(L_1, L_2)\Big\} =\\\noalign{\smallskip}
&=\dfrac{(K_2+L_2)\varrho^2}{2\,m_2}+(K_2-L_2)\varrho\ +\\\noalign{\smallskip}
&\qquad +\dfrac{(K_1+L_1)(1-m_2)^2+(K_2+L_2)m_1m_2}{2\,m_1}
\ea
\ee
with
\[
\alpha = \dfrac{1}{2}\left\{(L_2-K_2)+\dfrac{m_2}{\varrho\,m_1}\big[(1-m_2)(K_1+L_1)-m_1(K_2+L_2)\big]\right\} .
\]
Spherical and deviatoric components of optimal average stress in phases are given by
\be
\label{e321-15}
\ba{llll}
\mbox{in phase $1$:}
&S_1=\dfrac{1-m_2}{\sqrt{2}\,m_1}, 
&D_{11} = S_1,
&D_{12} = 0,\\\noalign{\smallskip} 
\mbox{in phase $2$:}
&S_2=\dfrac{m_2+\varrho}{\sqrt{2}\,m_2},
&D_{21}=\dfrac{m_2-\varrho}{\sqrt{2}\,m_2},
&D_{22} = 0 .
\ea
\ee

By substituting \eqref{e321-15} in \eqref{e31-6} and \eqref{e31-3} respectively we obtain sufficient optimality condition of $U_\mrm{tr}$ in region $\cC$ expressed in terms of stress fields in phases 1 and 2.

Region $\cC$ is represented in Fig.\ \ref{fig2} by a curvilinear triangle $P_1P_2P_3$. Its boundaries are determined by the following expressions
\[
\ba{lll}
\alpha < L_2 &\Rightarrow &m_1 > \psi_{\cA-\cC}(m_2,\varrho),\\\noalign{\smallskip}
\alpha > L_1 &\Rightarrow &m_1 < \psi_{\cC-\cE}(m_2,\varrho),\\\noalign{\smallskip}
m_1S_1\leq D_0 &\Rightarrow &\varrho\in [0,m_2]
\ea
\]
where
\be
\label{e321-17}
\ba{l}
\psi_{\cA-\cC}(m_2,\varrho)=\dfrac{m_2(1-m_2)(K_1+L_1)}{(m_2+\varrho)(K_2+L_2)},\\\noalign{\smallskip}
\psi_{\cC-\cE}(m_2,\varrho)=\dfrac{m_2(1-m_2)(K_1+L_1)}{(m_2+\varrho)(K_2+L_2)-2(L_2-L_1)\varrho}.
\ea
\ee


\subsubsection{Case of $\varrho\in [0,1]$ and $\alpha > L_2$: Region $\mathsf{A}$}
\label{sec322-regionA}

According to the discussion in Sec.\ \ref{sec31-fields}, item (IIa), set
\[
\Phi_i = m_i(K_i+L_i)\,S_i\2,\qquad i=1,2.
\]
Hence
\[
\ba{ll}
\Phi(\varrho,\alpha) = &\min\big\{\Phi_1 + \Phi_2 \big\}\\\noalign{\smallskip}
&\mbox{subject to:}\ m_1S_1+m_2S_2 = S_0
\ea
\]
and
\[
U_{\mrm{tr}}(\varrho) = \max\Big\{\Phi(\varrho,\alpha) - 2\,\varrho\,\alpha\ \Big|\ \alpha\geq L_2\Big\} .
\]
It is immediate that the function to be maximized monotonically decreases in $\alpha$ hence we set
$\alpha=L_2$. Results obtained in the remainder of this section correspond to the region of optimality $\cA$ in Fig.\ \ref{fig2}. 

\subparagraph*{Region $\cA$\ :} Proceeding analogously to previous cases we derive the lower estimate of stress energy in region $\cA$. It takes the form
\be
\label{e322-4}
U_{\mrm{tr}}(\varrho) = \dfrac{(1+\varrho)^2}{2}\,\dfrac{(K_1+L_1)(K_2+L_2)}{m_1(K_2+L_2)+m_2(K_1+L_1)} - 2\varrho L_2.
\ee

Optimal values of average spherical components of stresses read
\be
\label{e322-5}
\ba{ll}
\mbox{in phase $1$:}
&S_1=\dfrac{1+\varrho}{\sqrt{2}}\,\dfrac{K_2+L_2}{m_1(K_2+L_2)+m_2(K_1+L_1)} ,\\\noalign{\smallskip}
\mbox{in phase $2$:}
&S_2=\dfrac{1+\varrho}{\sqrt{2}}\,\dfrac{K_1+L_1}{m_1(K_2+L_2)+m_2(K_1+L_1)} .
\ea
\ee

Substituting $S_1$ and $S_2$ in \eqref{e31-6} and \eqref{e31-7} leads to sufficient optimality condition of $U_\mrm{tr}$ in region $\cA$ expressed in terms of stress fields in phases 1 and 2 respectively.

For fixed $K_i$, $L_i$, $i=1,2$, and arbitrary $m_2$, region $\cA$ is described by
\[
\ba{ll}
\cA=\Big\{(\varrho,m_1)\colon 
&0\leq m_1\leq\psi_{\cA-\cC}(m_2,\varrho)\ \mbox{if $\varrho\in[0,m_2]$},\\\noalign{\smallskip}
&0\leq m_1\leq\psi_{\cA-\cB}(m_2,\varrho)\ \mbox{if $\varrho\in[m_2,1]$} \Big\}
\ea
\]
see \eqref{e321-10} and \eqref{e321-17}. Curvilinear sides of $\cA$ are represented in Fig.\ \ref{fig2} by lines $P_1P_3$ and $P_3P_6$. Note that $\cA$ splits into $\mathsf{A_1}$ and $\mathsf{A_2}$ with the interface represented by a curve $P_3P_7$. This division is explained in Sec.\ \ref{sec423-regionAA'}.


\subsubsection{Case of $\varrho\in [-1,0]$ and $\alpha\in (-K_2, -K_1)$: Regions $\mathsf{B'}$ and $\mathsf{C'}$}
\label{sec323-regionB'C'}

According to the discussion in Sec. \ref{sec31-fields}, items (I) and (IIIa), set
\[
\ba{l}
\Phi_1 = m_1(K_1+L_1)\,\left(D_{11}\2 +D_{12}\2\right),\\\noalign{\smallskip}
\Phi_2 = m_2(K_2+\alpha)\,S_2\2 +m_2(L_2-\alpha)\left(D_{21}\2 +D_{22}\2\right) .
\ea
\]
It follows that
\be
\label{e323-2}
\ba{ll}
\Phi(\varrho,\alpha) = &\min\big\{ \Phi_1 + \Phi_2 \big\}\\\noalign{\smallskip}
&\mbox{subject to:}\ m_1S_1+m_2S_2 = S_0,\\\noalign{\smallskip}
&\phantom{subject to:}\ m_1D_{11}+m_2D_{21} = D_0,\\\noalign{\smallskip}
&\phantom{subject to:}\ m_1D_{12}+m_2D_{22} = 0,\\\noalign{\smallskip}
&\phantom{subject to:}\ S_1\2\leq D_{11}\2 + D_{12}\2,\\\noalign{\smallskip}
&\phantom{subject to:}\ S_2\2\leq D_{21}\2 + D_{22}\2.
\ea
\ee

From the KKT optimality conditions we conclude that: (i) $D_{12}=D_{22}=0$, (ii) $S_2$ minimizes $\Phi_1+\Phi_2$ if $S_1$ takes its maximal value. All constraints in \eqref{e323-2} are satisfied if we set
\[
S_1 = \min\left\{\dfrac{1}{m_1}S_0,\ D_{11}\right\} .
\]
Similarly to Sec.\ \ref{sec321-regionBC}, the case splits into two subcases. They correspond to regions of optimality $\cB'$ and $\cC'$ in Fig.\ \ref{fig2}. 

\subparagraph*{Region $\cB'$\ :} Assume that
\[
S_1 = \dfrac{1}{m_1}S_0\leq D_{11}.
\]
By this, $S_2=0 $ and \eqref{e323-2} transforms to
\[
\ba{ll}
\Phi(\varrho,\alpha) = &\min\Big\{m_1(K_1+L_1)\,D_{11}\2 + m_2(K_2+\alpha)\,D_{21}\2\Big\}\\\noalign{\smallskip}
&\mbox{subject to:}\quad m_1D_{11}+m_2D_{21} = D_0 .
\ea
\]

Proceeding analogously to previous sections we obtain
\be
\label{e323-6}
\ba{ll}
U_{\mrm{tr}}(\varrho) &= \max\Big\{\Phi(\varrho,\alpha) - 2\,\varrho\,\alpha\ \Big|\ \alpha\in(-K_2, -K_1)\Big\} =\\\noalign{\smallskip}
&= \dfrac{(1-\varrho-2\sqrt{-\varrho\,m_2})^2}{2\,m_1}(K_1+L_1) - 2\varrho L_2
\ea
\ee
with
\[
\alpha = \dfrac{1}{2}\,\dfrac{\sqrt{-\varrho\,m_2}(1-\varrho-2\sqrt{-\varrho\,m_2})}{\varrho\,m_1}(K_1+L_1)+L_2 .
\]

Optimal average values of spherical and deviatoric components of stresses in region $\cB'$ are thus given by 
\be
\label{e323-8}
\ba{llll}
\mbox{in phase $1$:}
&S_1 = \dfrac{1+\varrho}{\sqrt{2}\,m_1},
&D_{11} = \dfrac{1-\varrho-2\sqrt{-\varrho\,m_2}}{\sqrt{2}\,m_1},
&D_{12} = 0,\\\noalign{\smallskip}
\mbox{in phase $2$:}
&S_2 = 0,
&D_{21} = \sqrt{2}\,\dfrac{\sqrt{-\varrho\,m_2}}{m_2} , 
&D_{22} = 0.
\ea
\ee

Sufficient optimality condition of $U_\mrm{tr}$ in region $\cB'$ results from substituting $S_1$, $D_{11}$, $D_{12}$ in \eqref{e31-10} and \eqref{e31-3}.

Region $\cB'$ is represented by a curvilinear rectangle $P'_2P'_3P'_5P'_6$ in Fig.\ \ref{fig2}. Its boundaries are determined according to the following scheme
\[
\ba{lll}
\alpha > -K_2 &\Rightarrow &m_1 > \psi_{\cA'-\cB'}(m_2,\varrho),\\\noalign{\smallskip}
\alpha < -K_1 &\Rightarrow &m_1 < \psi_{\cB'-\cD'}(m_2,\varrho),\\\noalign{\smallskip}
m_1D_{11}\geq S_0 &\Rightarrow &\varrho\in [-1,-m_2]
\ea
\]
where
\be
\label{e323-10}
\ba{l}
\psi_{\cA'-\cB'}(m_2,\varrho) = -\dfrac{\sqrt{-\varrho\,m_2}(1-\varrho-2\sqrt{-\varrho\,m_2})}{2\,\varrho}\,\dfrac{K_1+L_1}{K_2+L_2},\\\noalign{\smallskip}
\psi_{\cB'-\cD'}(m_2,\varrho) = -\dfrac{\sqrt{-\varrho\,m_2}(1-\varrho-2\sqrt{-\varrho\,m_2})}{2\,\varrho}\,\dfrac{K_1+L_1}{K_1+L_2}.
\ea
\ee

\subparagraph*{Region $\cC'$\ :} Conversely to previous paragraph assume
\[
S_1 = D_{11}\leq \dfrac{1}{m_1}S_0,
\]
which results in
\[
\ba{ll}
\Phi(\varrho,\alpha) = &\min\Big\{m_1(K_1+L_1)\,D_{11}\2 +\\
&\phantom{\min} + m_2\big[(K_2+\alpha)\,S_2\2 + (L_2-\alpha)D_{21}\2\big]\Big\}\\\noalign{\smallskip}
&\mbox{subject to:}\ m_1D_{11}+m_2S_2 = S_0,\\\noalign{\smallskip}
&\phantom{subject to:}\ m_1D_{11}+m_2D_{21} = D_0 .
\ea
\]
It follows that the stress energy in region $\cC'$ is bounded from below by
\be
\label{e323-13}
\ba{ll}
U_{\mrm{tr}}(\varrho) &= \max\Big\{\Phi(\varrho,\alpha) - 2\,\varrho\,\alpha\ \Big|\ \alpha\in(-K_2, -K_1)\Big\} =\\\noalign{\smallskip}
&=\dfrac{(K_2+L_2)\varrho^2}{2\,m_2}+(K_2-L_2)\varrho\ +\\\noalign{\smallskip}
&\qquad +\dfrac{(K_1+L_1)(1-m_2)^2+(K_2+L_2)m_1m_2}{2\,m_1}
\ea
\ee
with
\[
\alpha = \dfrac{1}{2}\left\{(L_2-K_2)-\dfrac{m_2}{\varrho\,m_1}\big[(1-m_2)(K_1+L_1)-m_1(K_2+L_2)\big]\right\} .
\]
Spherical and deviatoric components of optimal average stress in phases are given by
\be
\label{e323-15}
\ba{llll}
\mbox{in phase $1$:}
&S_1=\dfrac{1-m_2}{\sqrt{2}\,m_1}, 
&D_{11} = S_1,
&D_{12} = 0,\\\noalign{\smallskip} 
\mbox{in phase $2$:}
&S_2=\dfrac{m_2+\varrho}{\sqrt{2}\,m_2},
&D_{21}=\dfrac{m_2-\varrho}{\sqrt{2}\,m_2},
&D_{22} = 0 .
\ea
\ee

Substituting $S_1$, $D_{11}$, $D_{12}$ in \eqref{e31-10} and $S_2$, $D_{21}$, $D_{22}$ in \eqref{e31-3} results in sufficient optimality condition of $U_\mrm{tr}$ in region $\cC'$.

Region $\cC'$ is represented in Fig.\ \ref{fig2} by a curvilinear triangle $P_1P'_2P'_3$. Its boundaries are determined by the following expressions
\[
\ba{lll}
\alpha > -K_2 &\Rightarrow &m_1 > \psi_{\cA'-\cC'}(m_2,\varrho),\\\noalign{\smallskip}
\alpha < -K_1 &\Rightarrow &m_1 < \psi_{\cC'-\cE}(m_2,\varrho),\\\noalign{\smallskip}
m_1D_{11}\leq S_0 &\Rightarrow &\varrho\in [-m_2, 0]
\ea
\]
where
\be
\label{e323-17}
\ba{l}
\psi_{\cA'-\cC'}(m_2,\varrho)=\dfrac{m_2(1-m_2)(K_1+L_1)}{(m_2-\varrho)(K_2+L_2)},\\\noalign{\smallskip}
\psi_{\cC'-\cE}(m_2,\varrho)=\dfrac{m_2(1-m_2)(K_1+L_1)}{(m_2+\varrho)(K_2+L_2)-2(L_2+K_1)\varrho}.
\ea
\ee


\subsubsection{Case of $\varrho\in [-1,0]$ and $\alpha < -K_2$: Region $\mathsf{A'}$}
\label{sec324-regionA'}

According to the discussion in Sec.\ \ref{sec31-fields}, item (IIIa), set
\[
\Phi_i = m_i(K_i+L_i)\,\left(D_{i1}\2 +D_{i2}\2\right),\qquad i=1,2.
\]
Hence
\[
\ba{ll}
\Phi(\varrho,\alpha) = &\min\big\{\Phi_1 + \Phi_2\}\\\noalign{\smallskip}
&\mbox{subject to:}\ m_1D_{11}+m_2D_{21} = D_0,\\\noalign{\smallskip}
&\phantom{subject to:}\ m_1D_{12}+m_2D_{22} = 0,
\ea
\]
and it is immediate that $D_{12}=D_{22}=0$. 

The estimate of the stress energy is determined as
\be
\label{e324-3}
U_{\mrm{tr}}(\varrho) = \max\Big\{\Phi(\varrho,\alpha) - 2\,\varrho\,\alpha\ \Big|\ \alpha\leq -K_2\Big\} .
\ee
Function to be maximized in \eqref{e324-3} monotonically decreases in $\alpha$ hence we set $\alpha=-K_2$. Results obtained in the remainder of this section correspond to the region of optimality $\cA'$ in Fig.\ \ref{fig2}. 

\subparagraph*{Region $\cA'$\ :} The lower estimate of stress energy in region $\cA'$ takes the form
\be
\label{e324-4}
U_{\mrm{tr}}(\varrho) = \dfrac{(1-\varrho)^2}{2}\,\dfrac{(K_1+L_1)(K_2+L_2)}{m_1(K_2+L_2)+m_2(K_1+L_1)} + 2\varrho K_2.
\ee

Optimal values of average deviatoric fields read
\be
\label{e324-5}
\ba{ll}
\mbox{in phase $1$:}
&D_{11}=\dfrac{1-\varrho}{\sqrt{2}}\,\dfrac{K_2+L_2}{m_1(K_2+L_2)+m_2(K_1+L_1)} ,\\\noalign{\smallskip}
\mbox{in phase $2$:}
&D_{21}=\dfrac{1-\varrho}{\sqrt{2}}\,\dfrac{K_1+L_1}{m_1(K_2+L_2)+m_2(K_1+L_1)} .
\ea
\ee

Considering \eqref{e324-5} in \eqref{e31-10} and \eqref{e31-11} leads to sufficient optimality condition of $U_\mrm{tr}$ in region $\cA'$ expressed in terms of stress fields in phases 1 and 2 respectively.

For fixed $K_i$, $L_i$, $i=1,2$, region $\cA'$ is described by
\[
\ba{ll}
\cA'=\Big\{(\varrho,m_1,m_2)\colon 
&0\leq m_1\leq\psi_{\cA'-\cC'}(m_2,\varrho)\ \mbox{if $\varrho\in[-m_2,0]$},\\\noalign{\smallskip}
&0\leq m_1\leq\psi_{\cA'-\cB'}(m_2,\varrho)\ \mbox{if $\varrho\in[-1,m_2]$} \Big\}
\ea
\]
see \eqref{e323-10} and \eqref{e323-17}. Curvilinear sides of $\cA'$ are represented in Fig.\ \ref{fig2} by lines $P_1P'_3$ and $P'_3P'_6$. Note that $\cA'$ splits into $\mathsf{A'_1}$ and $\mathsf{A'_2}$ with the interface represented by a curve $P'_3P'_7$. This division is explained in Sec.\ \ref{sec423-regionAA'}.


\subsubsection{Case of $\varrho\in [-1,1]$ and $\alpha\in(-K_1, L_1)$: Regions $\mathsf{D}$, $\mathsf{D'}$ and $\mathsf{E}$}
\label{sec325-regionDD'E}
The case of $\alpha\in (-K_1, L_1)$ is discussed in Sec.\ \ref{sec31-fields}, item (I). Both $\Phi_1$ and $\Phi_2$ are described by
\[
\Phi_i = m_i\big[(K_i+\alpha)\,S_i\2 + (L_i-\alpha)\big(D_{i1}\2 + D_{i2}\2\big)\big],\quad i=1,2 .
\]
For $\varrho\in[0,1]$ it follows that 
\be
\label{e325-2}
\ba{ll}
\Phi(\varrho,\alpha) = &\min\big\{\Phi_1 + \Phi_2\big\}\\
\noalign{\smallskip}
&\mbox{subject to:}\ m_1S_1+m_2S_2 = S_0,\\\noalign{\smallskip}
&\phantom{subject to:}\ m_1D_{11}+m_2D_{21} = D_0,\\\noalign{\smallskip}
&\phantom{subject to:}\ m_1D_{12}+m_2D_{22} = 0,\\\noalign{\smallskip}
&\phantom{subject to:}\ S_1\2\geq D_{11}\2 + D_{12}\2,\\\noalign{\smallskip}
&\phantom{subject to:}\ S_2\2\geq D_{21}\2 + D_{22}\2 .
\ea
\ee
In case of $\varrho\in[-1,0]$, last two constraints change into
\[
\ba{l}
S_1\2\leq D_{11}\2 + D_{12}\2,\\\noalign{\smallskip}
S_2\2\leq D_{21}\2 + D_{22}\2 .
\ea
\]

The KKT requirements are that $D_{12}=D_{22}=0$. Hence, in case of arbitrary $\varrho$, problem \eqref{e325-2} takes the form
\[
\ba{ll}
\Phi(\varrho,\alpha) = &\min\Big\{m_1\big[(K_1+\alpha)\,S_1\2 + (L_1-\alpha)\,D_{11}\2\big] +\\
&\phantom{\min} + m_2\big[(K_2+\alpha)\,S_2\2 + (L_2-\alpha)\,D_{21}\2\Big\}\\
\noalign{\smallskip}
&\mbox{subject to:}\ m_1S_1+m_2S_2 = S_0,\\\noalign{\smallskip}
&\phantom{subject to:}\ m_1D_{11}+m_2D_{21} = D_0 .
\ea
\]

Applying necessary optimality conditions leads to
\[
\ba{ll}
 S_1 = \dfrac{(K_2+\alpha)\,S_0}{m_1(K_2+\alpha)+m_2(K_1+\alpha)},
&D_{11} = \dfrac{(L_2-\alpha)\,D_0}{m_1(L_2-\alpha)+m_2(L_1-\alpha)},\\\noalign{\smallskip}
 S_2 = \dfrac{(K_1+\alpha)\,S_0}{m_1(K_2+\alpha)+m_2(K_1+\alpha)},
&D_{21} = \dfrac{(L_1-\alpha)\,D_0}{m_1(L_2-\alpha)+m_2(L_1-\alpha)}.
\ea
\]

Technically, determining $U_{\mrm{tr}}(\varrho)$ and optimal $\alpha\in(-K_1, L_1)$ requires similar algorithm to the one used in previous sections. Corresponding formulae are not presented here due to their complexity. The limiting cases of $\alpha=L_1$ or $\alpha=-K_1$ are dealt with separately in the sequel of this Section. As a result, three additional regions of optimality appear. Namely, region $\cE$ corresponds to $\alpha\in(-K_1, L_1)$ and regions $\cD$ and $\cD'$ are related to $\alpha=L_1$ and $\alpha=-K_1$ respectively.

\subparagraph*{Region $\cE$\ :} The problem is now to find the formulae for two curves which separate region $\cE$ from $\cD$ and $\cD'$. These curves are given by implicit functions $\psi_{\cD-\cE}(m_1,m_2,\varrho) = 0$ and $\psi_{\cD'-\cE}(m_1,m_2,\varrho) = 0$ where
\[
\ba{l}
\psi_{\cD-\cE}(m_1,m_2,\varrho) = \dfrac{\partial}{\partial\alpha}\Big[\Phi(\varrho,\alpha) - 2\,\varrho\,\alpha\Big]\bigg|_
{\alpha=L_1},\\\noalign{\smallskip}
\psi_{\cD'-\cE}(m_1,m_2,\varrho) = \dfrac{\partial}{\partial\alpha}\Big[\Phi(\varrho,\alpha) - 2\,\varrho\,\alpha\Big]\bigg|_
{\alpha=-K_1} .
\ea
\]
It follows that
\[
\ba{l}
\psi_{\cD-\cE}(m_1,m_2,\varrho) = \widehat{\psi}_{\cD-\cE}(m_1,m_2)(1+\varrho)^2 + \dfrac{2\,\varrho\,(1-m_1)}{m_1},\\\noalign{\smallskip}
\widehat{\psi}_{\cD-\cE}(m_1,m_2) =\\
\quad\quad = -\dfrac{m_2(K_1+L_1)\big[(m_1+m_2)(K_1+L_1)+2\,m_1(K_2-K_1)\big]}{2\,m_1\big[m_1(K_2+L_1)+m_2(K_1+L_1)\big]^2} ,
\ea
\]
and
\[
\ba{l}
\psi_{\cD'-\cE}(m_1,m_2,\varrho) = \widehat{\psi}_{\cD'-\cE}(m_1,m_2)(1-\varrho)^2 + \dfrac{2\,\varrho\,(1-m_1)}{m_1},\\\noalign{\smallskip}
\widehat{\psi}_{\cD'-\cE}(m_1,m_2) =\\
\quad\quad = \dfrac{m_2(K_1+L_1)\big[(m_1+m_2)(K_1+L_1)+2\,m_1(L_2-L_1)\big]}{2\,m_1\big[m_1(K_1+L_2)+m_2(K_1+L_1)\big]^2} .
\ea
\]

Region $\cE$ is bounded by the lines
\[
\ba{ll}
m_1 = 1-m_2,\\\noalign{\smallskip}
m_1 = \psi_{\cC'-\cE} &\mbox{if $\varrho\in[-1,0]$},\\\noalign{\smallskip}
m_1 = \psi_{\cC-\cE} &\mbox{if $\varrho\in[0,\ 1]$},\\\noalign{\smallskip}
\psi_{\cD'-\cE}(m_1,m_2,\varrho) = 0,\\\noalign{\smallskip}
\psi_{\cD-\cE}(m_1,m_2,\varrho) = 0,
\ea
\]
see \eqref{e321-17} and \eqref{e323-17}. Curvilinear sides of $\cE$ described above are respectively represented in Fig.\ \ref{fig2} by lines $P_1P'_3$, $P_1P_3$, $P'_2P'_4$ and $P_2P_4$.

\subparagraph*{Region $\cD$\ :} Assuming $\alpha = L_1$ and repeating the discussion for region $\cE$ or region $\cB$ (Sec.\ \ref{sec321-regionBC}) we conclude that the stress energy estimation reads
\be
\label{e325-10}
U_{\mrm{tr}}(\varrho) = \dfrac{(1+\varrho)^2}{2}\,\dfrac{(K_1+L_1)(K_2+L_1)}{m_1(K_2+L_1)+m_2(K_1+L_1)} - 2\varrho L_1.
\ee
and optimal average fields in $\cD$ are given by
\be
\label{e325-11}
\ba{lll}
\mbox{in phase $1$:}\\\noalign{\smallskip}
S_1 = \dfrac{(K_2+L_1)(1+\varrho)}{\sqrt{2}\,\big[m_1(K_2+L_1)+m_2(K_1+L_1)\big]},
&D_{11} = \dfrac{1-\varrho}{\sqrt{2}\,m_1},
&D_{12} = 0,\\\noalign{\smallskip}
\mbox{in phase $2$:}\\\noalign{\smallskip}
S_2 = \dfrac{(K_1+L_1)(1+\varrho)}{\sqrt{2}\,\big[m_1(K_2+L_1)+m_2(K_1+L_1)\big]},
&D_{21} = 0, 
&D_{22} = 0 .
\ea
\ee

Region $\cD$ is represented in Fig.\ \ref{fig2} by an area bounded by two straight lines: (i) $\varrho=1$, (ii) $m_1=1-m_2$ and two curves: (i) $\psi_{\cD-\cE}(m_1,m_2,\varrho) = 0$ (line $P_2P_4$), (ii) $m_1 = \psi_{\cB-\cD}$ (line $P_2P_5$).

\subparagraph*{Region $\cD'$\ :} Assuming $\alpha = -K_1$ and repeating the discussion for region $\cE$ or region $\cB'$ (Sec.\ \ref{sec323-regionB'C'}) we conclude that the stress energy estimation reads
\be
\label{e325-12}
U_{\mrm{tr}}(\varrho) = \dfrac{(1-\varrho)^2}{2}\,\dfrac{(K_1+L_1)(K_1+L_2)}{m_1(K_1+L_2)+m_2(K_1+L_1)} + 2\varrho K_1.
\ee
and optimal average fields in $\cD'$ are given by
\be
\label{e325-13}
\ba{lll}
\mbox{in phase $1$:}\\\noalign{\smallskip}
S_1 = \dfrac{1+\varrho}{\sqrt{2}\,m_1},
&D_{11} = \dfrac{(K_1+L_2)(1-\varrho)}{\sqrt{2}\,\big[m_1(K_1+L_2)+m_2(K_1+L_1)\big]}
&D_{12} = 0,\\\noalign{\smallskip}
\mbox{in phase $2$:}\\\noalign{\smallskip}
S_2 = 0,
&D_{21} = \dfrac{(K_1+L_1)(1-\varrho)}{\sqrt{2}\,\big[m_1(K_1+L_2)+m_2(K_1+L_1)\big]}, 
&D_{22} = 0.
\ea
\ee

Region $\cD'$ is represented in Fig.\ \ref{fig2} by an area bounded by two straight lines: (i) $\varrho=-1$, (ii) $m_1=1-m_2$ and two curves: (i) $\psi_{\cD'-\cE}(m_1,m_2,\varrho) = 0$ (line $P'_2P'_4$), (ii) $m_1 = \psi_{\cB'-\cD'}$ (line $P'_2P'_5$).

Sufficient optimality condition of $U_\mrm{tr}$ in region $\cD$, see \eqref{e325-10}, results from substituting \eqref{e325-13} in \eqref{e31-7} and \eqref{e31-3}. Similarly, considering \eqref{e325-13} in \eqref{e31-11} and \eqref{e31-3} leads to sufficient optimality condition of $U_\mrm{tr}$ given by \eqref{e325-12} in region $\cD'$.


\subsection{Bounds on effective isotropic properties}
\label{sec33-Gclosure}

Making use of \eqref{e22-6} allows for calculating bounds on effective isotropic properties in each optimality region where $U_\ast(\varrho)$ is determined. From the results obtained in the preceeding section and by assuming that $U_{\mrm{tr}}(\varrho) = U_\ast(\varrho)$ it follows that formulae for $K_\ast(\varrho)$ and $L_\ast(\varrho)$ can be derived in any region except $\cE$. Recall that $K_\ast(\varrho)$ and $L_\ast(\varrho)$ are related to $\partial G_mA$ only if optimal stress fields predicted in Sec.\ \ref{sec31-fields} are statically admissible. In Sec.\ \ref{sec4-laminates} we prove that this is the case for high-porosity regions. We conjecture the same property for $\cD$ and $\cD'$, see the discussion in Sec.\ \ref{sec5-remarks}.

\subparagraph*{Region $\cA$\ :}
\be
\label{e33-1}
K_\ast(\varrho) = \left(\dfrac{m_1}{K_1+L_1}+\dfrac{m_2}{K_2+L_2}\right)^{-1}-L_2,\qquad L_\ast(\varrho) = L_2 .
\ee

\subparagraph*{Region $\cA'$\ :}
\be
\label{e33-2}
K_\ast(\varrho) = K_2,\qquad L_\ast(\varrho) = \left(\dfrac{m_1}{K_1+L_1}+\dfrac{m_2}{K_2+L_2}\right)^{-1}-K_2 .
\ee

\subparagraph*{Region $\cB$\ :}
\be
\label{e33-3}
\ba{l}
K_\ast(\varrho) = K_2-\dfrac{\big[(1+\varrho)\sqrt{\varrho\,m_2}-2\,\varrho\big]\big[1+\varrho-2\sqrt{\varrho\,m_2}\big]}{2\,m_1\,\varrho(1+\varrho)}(K_1+L_1),\\\noalign{\smallskip}
L_\ast(\varrho) = \dfrac{\sqrt{\varrho\,m_2}(1+\varrho-2\sqrt{\varrho\,m_2})}{2\,m_1\,\varrho}(K_1+L_1)-K_2 .
\ea
\ee

\subparagraph*{Region $\cB'$\ :}
\be
\label{e33-4}
\ba{l}
K_\ast(\varrho) = -\dfrac{\sqrt{-\varrho\,m_2}(1-\varrho-2\sqrt{-\varrho\,m_2})}{2\,m_1\,\varrho}(K_1+L_1)-L_2,\\\noalign{\smallskip}
L_\ast(\varrho) = L_2+\dfrac{\big[(1-\varrho)\sqrt{-\varrho\,m_2}+2\,\varrho\big]\big[1-\varrho-2\sqrt{-\varrho\,m_2}\big]}{2\,m_1\,\varrho(1-\varrho)}(K_1+L_1) .
\ea
\ee
Note that $\varrho<0$ in $\cB'$.

\subparagraph*{Regions $\cC$ and $\cC'$\ :}
\be
\label{e33-5}
\ba{l}
K_\ast(\varrho) = \dfrac{1}{2}\left[(K_2-L_2)+\dfrac{(1-m_2)^2}{m_1(1+\varrho)}(K_1+L_1)+\dfrac{m_2\2+\varrho}{m_2(1+\varrho)}(K_2+L_2)\right],\\\noalign{\smallskip}
L_\ast(\varrho) = \dfrac{1}{2}\left[\dfrac{(1-m_2)^2}{m_1(1-\varrho)}(K_1+L_1)+\dfrac{m_2\2-\varrho}{m_2(1-\varrho)}(K_2+L_2)-(K_2-L_2)\right] .
\ea
\ee

\subparagraph*{Region $\cD$\ :}
\be
\label{e33-6}
K_\ast(\varrho) = \left(\dfrac{m_1}{K_1+L_1}+\dfrac{m_2}{K_2+L_1}\right)^{-1}-L_1,\qquad L_\ast(\varrho) = L_1 .
\ee

\subparagraph*{Region $\cD'$\ :}
\be
\label{e33-7}
K_\ast(\varrho) = K_1,\qquad L_\ast(\varrho) = \left(\dfrac{m_1}{K_1+L_1}+\dfrac{m_2}{K_1+L_2}\right)^{-1}-K_1 .
\ee
\begin{figure}[H]
\centering
\includegraphics[width=90mm]{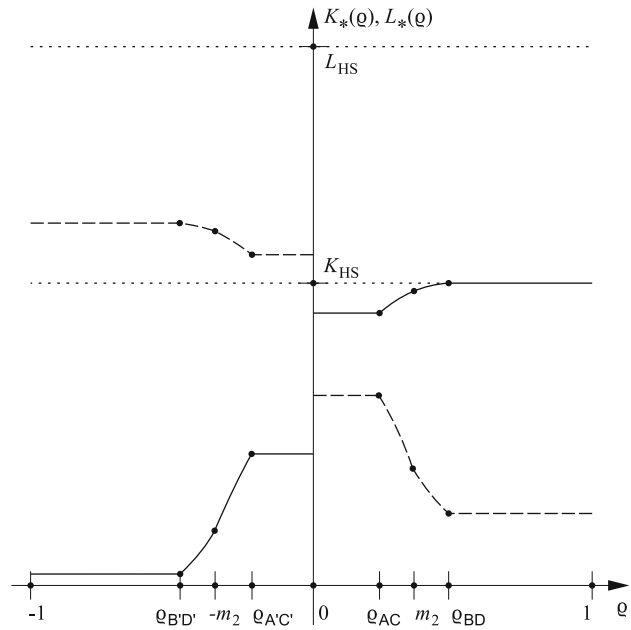}
\caption{Comparison of optimal bounds $K_\ast(\varrho)$ (solid lines), $L_\ast(\varrho)$ (dashed lines) and the Hashin-Shtrikman estimates $K_{\mrm{HS}}$, $L_{\mrm{HS}}$ (dotted lines). Values of functions are calculated for $m_1=0.17$, $m_2=0.35$ and $K_1=1$, $L_1=2$, $K_2=3$, $L_2=4$. Symbols $\varrho_{\mathsf{AC}}$, $\varrho_{\mathsf{BD}}$, $\varrho_{\mathsf{A'C'}}$, $\varrho_{\mathsf{B'D'}}$ refer to the anisotropy level of $\tau_0$ at the interfaces between respective regions; $\varrho=0$ at the interface between $\mathsf{A}$ and $\mathsf{A'}$, $\varrho=m_2$ at the interface between $\mathsf{C}$ and $\mathsf{B}$, $\varrho=-m_2$ at the interface between $\mathsf{C'}$ and $\mathsf{B'}$. \label{fig3}}
\end{figure}
Figure \ref{fig3} illustrates the comparison of functions $K_\ast(\varrho)$ and $L_\ast(\varrho)$ representing coupled lower bounds on isotropic properties of a three-phase composite in different regions with the Hashin-Shtrikman uncoupled bounds
\[
\ba{ll}
K_{\mrm{HS}} = \left(\dfrac{m_1}{K_1+\alpha_K}+\dfrac{m_2}{K_2+\alpha_K}\right)^{-1}-\alpha_K, &\alpha_K = L_1,\\\noalign{\smallskip}
L_{\mrm{HS}} = \left(\dfrac{m_1}{L_1+\alpha_L}+\dfrac{m_2}{L_2+\alpha_L}\right)^{-1}-\alpha_L, &\alpha_L = 2\,K_1+L_1 .
\ea
\]
Estimates $K_{\mrm{HS}}$, $L_{\mrm{HS}}$ are independent of $\varrho\in [-1,1]$, as they do not incorporate an information on the anisotropy of $\tau_0$. 

Note that $K_\ast(\varrho)\leq K_{\mrm{HS}}$ for all $\varrho\in [-1,1]$ and $K_\ast(\varrho)=K_{\mrm{HS}}$ in region $\cD$ while the inequality $L_\ast(\varrho) < L_{\mrm{HS}}$ is slack in all regions, see the discussion in Sec.\ \ref{sec5-remarks}.

%% file: 3mat-acgd-revised-sec4.tex
In this Section we show that optimal relaxed stress fields determined in regions $\cA, \cB, \cC$ and $\cA', \cB', \cC'$ coincide with statically admissible stress fields $\tau\in\Sigma$. The task is two-fold. First, we make use of the differential constraint $\mrm{div}\tau = 0$ in deriving additional requirements on $\tau\in\Sigma_{\mrm{rel}}$. Next, we show that these requirements are fulfilled in certain microstructures, so-called laminates of high rank.

\subsection{Compatibility of stresses on phase interfaces and average stresses in rank-one laminates}
\label{sec41-compatib}

In calculations of optimal $\tau\in\Sigma_{\mrm{rel}}$, the differential constraint $\mrm{div}\tau = 0$ in $Y$ (equilibrium equation) is neglected. Consequently, energy-minimizing stress fields are determined in each phase independently. It follows that components of optimal relaxed fields may be incompatible with $\mrm{div}\tau = 0$ on material interfaces which in turn means that $\tau\notin\Sigma$. 

Suppose that two materials meet in a given microstructure at a line $\Gamma$ and let $n$ and $t$ denote a normal and tangent to $\Gamma$. In the sequel we consider microstructures where phases are arranged in layers hence $\Gamma$ takes a form of a straight line. Moreover, we assume that stress field in each layer is constant. By this we claim that if a given non-degenerate phase $Y_i$, $i=1,2$, is distributed in $p$ layers $Y_{i,1}$, $Y_{i,2}$, $\ldots$, $Y_{i,p}$, then optimal $\tau$ is \emph{layer-wise constant} in $Y_i$. It follows that if $p=1$ then $\tau$ is constant in entire $Y_i$. Equlibrium equation is thus fulfilled identically in each phase.

Constraint $\mrm{div}\tau = 0$ requires that $[[\tau\,n]]_\Gamma = 0$ where $[[\,\cdot\,]]_\Gamma$ denotes a jump on the interface $\Gamma$ between layers of materials. If we set $\tau_m$, $m=1,2$, for constant fields on both sides of $\Gamma$ then the jump condition may be rewritten in a form
\[
(\tau_1-\tau_2):(n\otimes n) = (\tau_1-\tau_2):(n\otimes t) = (\tau_1-\tau_2):(t\otimes n) = 0 .
\]
It follows that stress fields with $(\tau_1-\tau_2):(t\otimes t)\neq 0$ are compatible with the equlibrium constraint hence statically admissible in $Y$, see \citep[Sec.\ 14.2.2]{Che00} for full discussion of this topic. 
\begin{figure}[H]
\centering
\includegraphics[width=90mm]{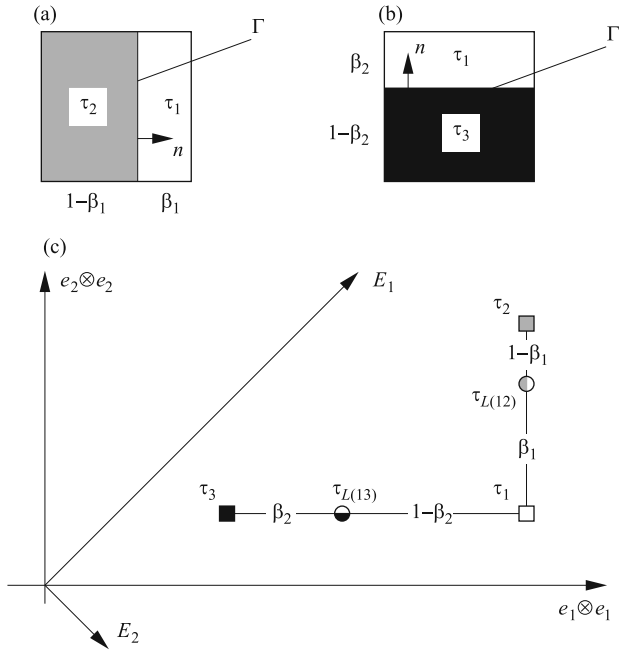}
\caption{Rank-one connectivity of stress fields $\tau_m$, $m=1,2,3$ such that $\tau_m:(e_1\otimes e_2)=0$: (a) simple laminate $L(12)$ with phases taken in proportions $\beta_1$, $1-\beta_1$, a normal to the interface $\Gamma$ given by $n=e_1$ and stress fields $\tau_1$ and $\tau_2$; (b) simple laminate $L(13)$ with phases taken in proportions $\beta_2$, $1-\beta_2$, a normal to the interface $\Gamma$ given by $n=e_2$ and constant stress fields $\tau_1$ and $\tau_3$; (c) graphical interpretation of compatibility conditions $(\tau_1-\tau_2):(e_1\otimes e_1) = 0$, $(\tau_1-\tau_3):(e_2\otimes e_2) = 0$ and average fields $\tau_{L(12)}$ in laminate $L(12)$, $\tau_{L(13)}$ in laminate $L(13)$. Vectors $E_1$, $E_2$ are defined in \eqref{e21-1a}. \label{fig4}}
\end{figure}
Here we discuss stresses $\tau_A$, $\tau_B$ in two materials $A$ and $B$, arranged in a rank-one laminate $L(AB)$. Compatibility of stress fields in $L(AB)$ is also referred to as \emph{rank-one connectivity at $\Gamma$}. Let $\tau_A$ and $\tau_B$ denote rank-one connected stress fields in materials layered in proportions $\beta$ and $1-\beta$ respectively. Resulting average field takes a value $\tau_{L(AB)} = \beta\,\tau_A +(1-\beta)\,\tau_B$. Examples of rank-one connected stress fields and their average values in simple laminates are sketched in Fig.\ \ref{fig4}. High-rank laminates are constructed by repeated rank-one layering scheme under the assumption that the materials resulting from previous laminations are homogeneous. These type of structures are considered in the subsequent section.

\subsection{Optimal high-rank laminates}
\label{sec42-structures}

\subsubsection{Regions $\cC$ and $\cC'$}
\label{sec421-regionCC'}

\begin{figure}[H]
\centering
\includegraphics[width=90mm]{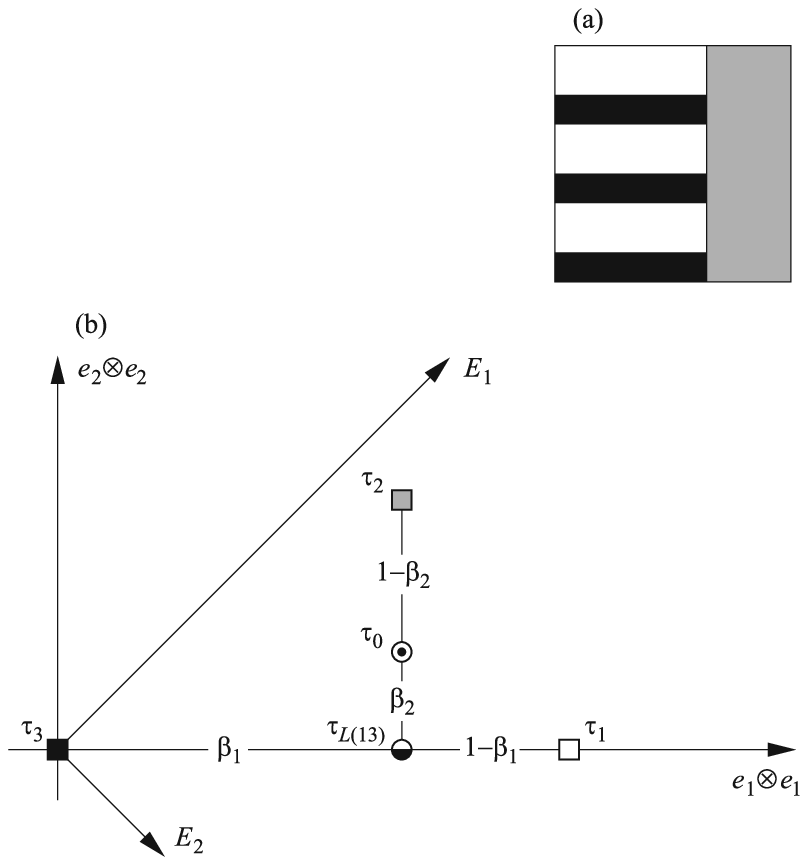}
\caption{Optimal microstructure in regions $\cC$ and $\cC'$: (a) layout of materials; (b) layering scheme leading to $\tau_0 = \tau_{L(13,2)}$ for region $\cC$ (a mirror image with respect to $e_1\otimes e_1$ results in the scheme for region $\cC'$). Stress fields in layers of strong and weak materials are represented by white and grey squares respectively, $\tau = 0$ in void is represented by black square. Circles represent stress fields in laminates.\label{fig5}}
\end{figure}

\subparagraph*{Region $\cC$\ :} Continuing the discussion in Sec.\ \ref{sec321-regionBC} one may notice that the assumption $S_1=D_{11}$ enforces $\theta(y)=0$ a.e. in $Y_1$. Consequently, optimal stress field $\tau\in\Sigma_{\mrm{rel}}$ is constant in $Y_1$. We thus calculate
\be
\label{e421-1}
\ba{ll}
\tau_1 = \dfrac{1-m_2}{m_1}\,e_1\otimes e_1 &\mbox{a.e.\ in}\ Y_1,\\\noalign{\medskip}
\tau_2 = 1\,e_1\otimes e_1 + \dfrac{\varrho}{m_2}\,e_2\otimes e_2 &\mbox{a.e.\ in}\ Y_2 .
\ea 
\ee
In order to prove statical admissibility of \eqref{e421-1} we check the compatibility of stresses on material interfaces in a $L(13,2)$ laminate, see Fig.\ \ref{fig5}.

\noindent\underline{Outline of the layering scheme}:
\begin{enumerate}
\item Substructure $L(13)$ is formed: phase 1 and void are laminated with $n_1 = e_2$ and volume fractions $\beta_1$, $1-\beta_1$ respectively. Homogenized stress field in $L(13)$ is given by $\tau_{L(13)} = \beta_1\,\tau_1$.

\item Final structure $L(13,2)$ is formed: phase 2 and $L(13)$ are laminated with $n_2=e_1$ and volume fractions $\beta_2$, $1-\beta_2$ respectively. Fields $\tau_{L(13)}$ and $\tau_2$ are rank-one connected if $(\tau_{L(13)} - \tau_2):(e_1\otimes e_1) = 0$. Stress field in the final structure $\tau_{L(13,2)}=\beta_2\,\tau_2 + (1-\beta_2)\,\tau_{L(13)}$ satisfies $\tau_{L(13,2)}=\tau_0$.
\end{enumerate}

\noindent\underline{Parameters of optimal laminate}:
Compatibility conditions reduce to
\be
\label{e421-2}
\ba{lll}
\tau_{L(13)}:(e_1\otimes e_1) = 1 &\Rightarrow &\beta_1 = \dfrac{m_1}{1-m_2}\in [0,1],\\\noalign{\smallskip}
\tau_{L(13,2)}:(e_2\otimes e_2) = \varrho &\Rightarrow &\beta_2 = m_2\in [0,1]
\ea
\ee
and it is immediate that the constraints on volume fractions of phases in $Y$ given by
\[
\beta_1(1-\beta_2) = m_1,\quad \beta_2 = m_2
\]
are satisfied identically.

\subparagraph*{Region $\cC'$\ :} In addition to the considerations in Sec.\ \ref{sec323-regionB'C'} and due to assumed $S_1 = D_{11}$ we set $s(y)=d_1(y)=S_1$ a.e. in $Y_1$. Further discussion reduces to the one presented above with $\varrho\in[-1,0]$ taken into account. Spherical and deviatoric components of average stress in each phase are given by the same formulae in both regions $\cC$ and $\cC'$, see \eqref{e321-15} and \eqref{e323-15}. Consequently, stress field in laminate $L(13,2)$ fulfills the sufficient optimality condition also in region $\cC'$ with phase volume fractions given by \eqref{e421-2}. Layout of materials and scheme of layering corresponding to region $\cC'$ are sketched in Fig.\ \ref{fig5}.


\subsubsection{Regions $\cB$ and $\cB'$}
\label{sec422-regionBB'}

\subparagraph*{Region $\cB$\ :} Here we continue the discussion in Sec.\ \ref{sec321-regionBC} with the assumption $D_{11} < S_1$. Function $\theta(y)$, $y\in Y_1$, may vary in $Y_1$ hence the stress field in material 1 are rank-one connected with zero stress in void if $\theta(y)=0$ or $\theta(y)=\pi$. Taking this into consideration we subdivide phase 1 into two layers, i.e.\ we set $Y_1=Y_{1,1}+Y_{1,2}$. Formulae for stresses read
\be
\label{e422-1}
\ba{ll}
\tau_{1,1} = \dfrac{1+\varrho-2\,\sqrt{\varrho\,m_2}}{m_1}\,e_1\otimes e_1 &\mbox{a.e.\ in } Y_{1,1},\\\noalign{\smallskip}
\tau_{1,2} = \dfrac{1+\varrho-2\,\sqrt{\varrho\,m_2}}{m_1}\,e_2\otimes e_2 &\mbox{a.e.\ in } Y_{1,2},\\\noalign{\smallskip}
\tau_2 = \dfrac{\sqrt{\varrho\,m_2}}{m_2}(e_1\otimes e_1 + e_2\otimes e_2) &\mbox{a.e.\ in } Y_2.
\ea
\ee

\begin{figure}[H]
\centering
\includegraphics[width=90mm]{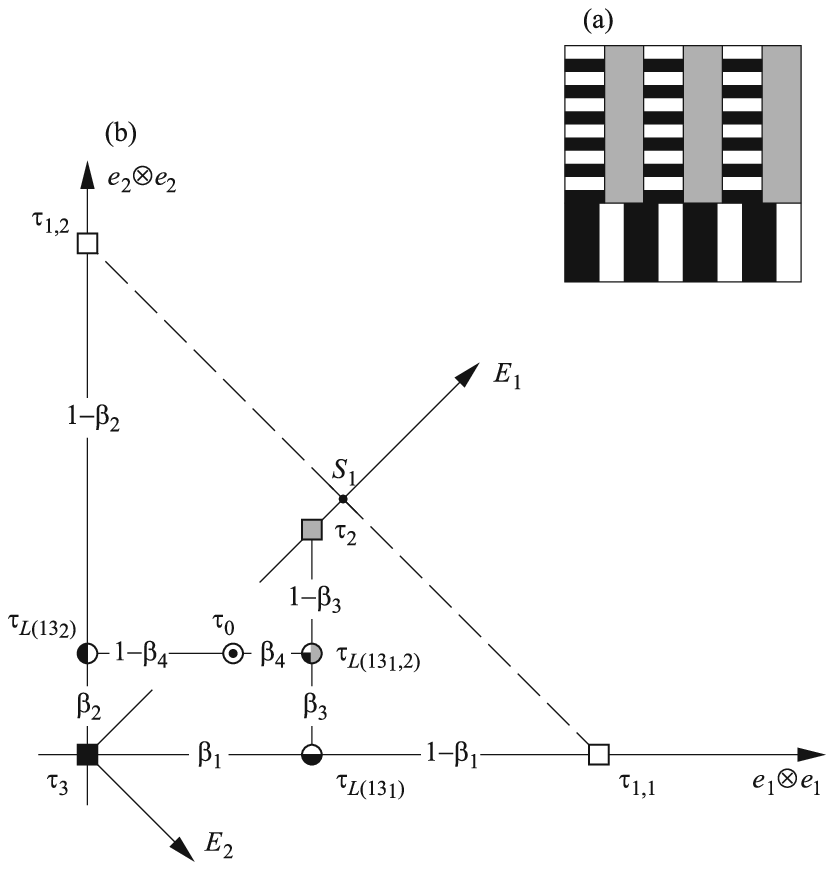}
\caption{Optimal microstructure in regions $\cB$ and $\cB'$: (a) layout of materials; (b) layering scheme leading to $\tau_0 = \tau_{L(13_1,2,13_2)}$ for region $\cB$ (a mirror image with respect to $e_1\otimes e_1$ results in the scheme for region $\cB'$). Stress fields in layers of strong and weak materials are represented by white and grey squares respectively, $\tau = 0$ in void is represented by black square. Circles represent stress fields in laminates.\label{fig6}}
\end{figure}

Next, we prove statical admissibility of \eqref{e422-1} by checking rank-one connectivity of stress fields in a $L(13_1,2,13_2)$ laminate, see Fig.\ \ref{fig6}.

\noindent\underline{Outline of the layering scheme}:
\begin{enumerate}
\item Substructures $L(13_1)$ and $L(13_2)$ are formed. In $L(13_1)$, the first layer of phase 1 (field $\tau_{1,1}$) and void are laminated with $n_1 = e_2$ and volume fractions $\beta_1$, $1-\beta_1$ respectively. Stress field in $L(13_1)$ is given by $\tau_{L(13_1)} = \beta_1\,\tau_{1,1}$. In $L(13_2)$, the second layer of phase 1 (field $\tau_{1,2}$) and void are laminated with $n_2 = e_1$ and volume fractions $\beta_2$, $1-\beta_2$. Stress field in $L(13_2)$ reads $\tau_{L(13_2)} = \beta_2\,\tau_{1,2}$.

\item Substructure $L(13_1,2)$ is formed: phase 2 and $L(13_1)$ are laminated with $n_3=e_1$ and volume fractions $\beta_3$, $1-\beta_3$. Fields $\tau_{L(13_1)}$ and $\tau_2$ are rank-one connected if $(\tau_{L(13_1)}-\tau_2):(e_1\otimes e_1) = 0$. Stress field in the substructure is given by $\tau_{L(13_1,2)}=\beta_3\,\tau_2 + (1-\beta_3)\,\tau_{L(13_1)}$.

\item Final structure $L(13_1,2,13_2)$ is formed: laminates $L(13_2)$ and $L(13_1,2)$ are layered with $n_4=e_2$ and volume fractions $\beta_4$ and $1-\beta_4$. Rank-one connection between stress fields holds if $(\tau_{L(13_2)}-\tau_{L(13_1,2)}):(e_2\otimes e_2) = 0$. Formulae $\tau_{L(13_1,2,13_2)} = \beta_4\,\tau_{0,2}+(1-\beta_4)\tau_{0,3}$ and $\tau_{L(13_1,2,13_2)}=\tau_0$ link the fields in substructures with the average stress tensor.
\end{enumerate}

\noindent\underline{Parameters of optimal laminate}:
Collecting the combatibility conditions we get the following
\[
\ba{lll}
(\tau_{L(13_1)}-\tau_2):(e_1\otimes e_1)=0 &\Rightarrow &\beta_1 = \dfrac{S_2}{2\,S_1}\ ,\\\noalign{\smallskip}
(\tau_{L(13_2)}-\tau_{L(13_1,2)}):(e_2\otimes e_2) = 0 &\Rightarrow &\beta_2 = \dfrac{\beta_3\,S_2}{2\,S_1}
\ ,\\\noalign{\smallskip}
\tau_{L(13_1,2,13_2)}:(e_2\otimes e_2) = \varrho &\Rightarrow &\beta_3 = \dfrac{\sqrt{2}\,\varrho}{S_2}\ ,\\\noalign{\smallskip}
\tau_{L(13_1,2,13_2)}:(e_1\otimes e_1) = 1 &\Rightarrow &\beta_4 = 1 - \dfrac{\sqrt{2}}{S_2} .
\ea
\]
Parameters $\beta_i\in[0,1]$, $i=1,\ldots,4$, as $\sqrt{2}\leq S_2\leq S_1$ in entire region $\cB$. Indeed, it is a matter of straightforward calculations to check that the first inequality immediately follows due to $\varrho\geq m_2$ in $\cB$ and the second one reduces to
\[
m_1\leq \psi_{\cB-\cD}(\varrho,m_2)\,\dfrac{K_1+L_2}{K_1+L_1}
\]
by substituting relevant formulae from \eqref{e321-8} and \eqref{e321-10}.

Constraints on volume fractions of phases in $Y$ are satisfied if $|Y_{1,1}|+|Y_{1,2}| = m_1$ and $|Y_2| = m_2$ or, equivalently,
\[
\ba{l}
(1-\beta_3)(1-\beta_4)\beta_1 + \beta_4\,\beta_2 = m_1\ ,\\\noalign{\smallskip}
(1-\beta_4)\beta_3 = m_2
\ea
\]
from which we have
\be
\label{e422-4}
\ba{ll}
\beta_1 = \dfrac{m_1\,\sqrt{\varrho\,m_2}}{m_2(1+\varrho-2\,\sqrt{\varrho\,m_2})}\ , &\beta_2 = \dfrac{m_1\,\varrho}{1+\varrho-2\,\sqrt{\varrho\,m_2}}\ ,\\\noalign{\smallskip}
\beta_3 = \sqrt{\varrho\,m_2}\ , &\beta_4 = 1-\sqrt{\dfrac{m_2}{\varrho}}\ .
\ea
\ee
Note that $L(13_1,2,13_2)$ (optimal in region $\cB$) morphs into $L(13,2)$ (optimal in $\cC$) at the boundary between regions. This is concluded by setting $\varrho = m_2$ in $\beta_1,\ldots,\beta_4$ above. 

\subparagraph*{Region $\cB'$\ :} From the discussion in Sec.\ \ref{sec323-regionB'C'} and the assumption that $S_1 < D_{11}$ it follows that phase 1 is subdivided into two layers such that $Y_1=Y_{1,1}+Y_{1,2}$ with $s(y) = D_{11}$ a.e.\ in $Y_{1,1}$ and $s(y) = - D_{11}$ a.e.\ in $Y_{1,2}$. In this way, stress fields are given by
\[
\ba{ll}
\tau_{1,1} = \dfrac{1-\varrho-2\,\sqrt{-\varrho\,m_2}}{m_1}\,e_1\otimes e_1 &\mbox{a.e.\ in } Y_{1,1},\\\noalign{\smallskip}
\tau_{1,2} = -\dfrac{1-\varrho-2\,\sqrt{-\varrho\,m_2}}{m_1}\,e_2\otimes e_2 &\mbox{a.e.\ in } Y_{1,2},\\\noalign{\smallskip}
\tau_2 = \dfrac{\sqrt{-\varrho\,m_2}}{m_2}(e_1\otimes e_1 - e_2\otimes e_2) &\mbox{a.e.\ in } Y_2.
\ea
\]

Thus described stress field in laminate $L(13_1,2,13_2)$ is statically admissible in region $\cB'$. Phase volume fractions are given by \eqref{e422-4} with $\varrho$ replaced with $-\varrho$. The details of calculations are omitted here as they follow the pattern presented above. Layout of materials and scheme of layering corresponding to region $\cB'$ are sketched in Fig.\ \ref{fig6}.


\subsubsection{Regions $\cA$ and $\cA'$}
\label{sec423-regionAA'}

Region $\cA$ splits into two subregions $\mathsf{A_1}$, $\mathsf{A_2}$ with different optimal microstructures. 

\begin{figure}[H]
\centering
\includegraphics[width=90mm]{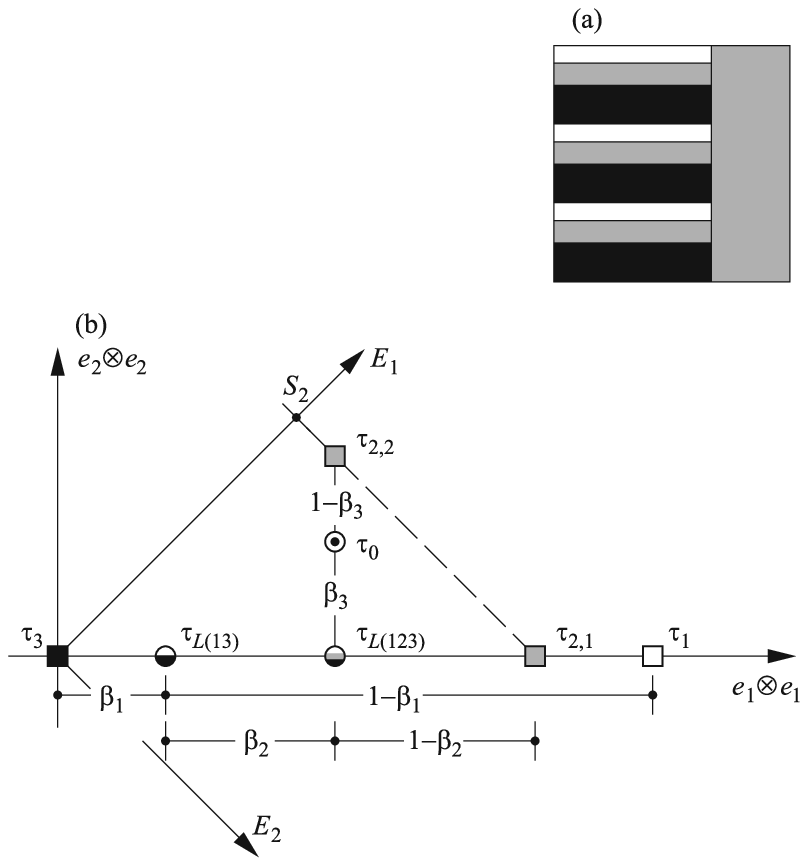}
\caption{Optimal microstructure in regions $\mathsf{A_1}$ and $\mathsf{A_1'}$: (a) layout of materials; (b) layering scheme leading to $\tau_0 = \tau_{L(132,2)}$ for region $\mathsf{A_1}$ (a mirror image with respect to $e_1\otimes e_1$ results in the scheme for region $\mathsf{A_1'}$). Stress fields in layers of strong and weak materials are represented by white and grey squares respectively, $\tau = 0$ in void is represented by black square. Circles represent stress fields in laminates.\label{fig7}}
\end{figure}

\subparagraph*{Subregion $\mathsf{A_1}$\ :}
We use the results of Sec.\ \ref{sec322-regionA} to prove that the stress fields
\[
\ba{ll}
\tau_1 = \sqrt{2}\,S_1\,e_1\otimes e_1 &\mbox{a.e.\ in } Y_1,\\\noalign{\medskip}
\tau_{2,1} = \sqrt{2}\,S_2\,e_1\otimes e_1 &\mbox{a.e.\ in } Y_{2,1},\\\noalign{\smallskip}
\tau_{2,2} = \dfrac{S_2+f_1}{\sqrt{2}}e_1\otimes e_1 + \dfrac{S_2-f_1}{\sqrt{2}}e_2\otimes e_2 &\mbox{a.e.\ in } Y_{2,2}
\ea
\]
are statically admissible in laminate $L(123,2)$, see Fig.\ \ref{fig7}. Phase 2 is thus subdivided into layers $Y_{2,1}$ and $Y_{2,2}$. In the sequel we assume $\tau_{2,2}:(e_1\otimes e_1)=1$ from which it follows that $f_1=\sqrt{2}-S_2$. Values of $S_1$ and $S_2$ are given by \eqref{e322-5}.

\noindent\underline{Outline of the layering scheme}:
\begin{enumerate}
\item Substructure $L(123)$ is formed in two steps: (i) phase 1 and void are layered with $n_1 = e_2$ and volume fractions $\beta_1$ and $1-\beta_1$ (in this way $L(13)$ with $\tau_{0,1} = \beta_1\,\tau_1$ is obtained), (ii) first layer of phase 2 and $L(13)$ are laminated in the same direction with volume fractions $\beta_2$ and $1-\beta_2$; this leads to $L(123)$ with $\tau_{L(123)} = \beta_2\,\tau_{2,1} + (1-\beta_2)\,\beta_1\,\tau_1$.

\item Final structure $L(123,2)$ is formed: second layer of phase 2 (field $\tau_{2,2}$) and $L(123)$ are laminated with $n_3 = e_1$ and volume fractions $\beta_3$ and $1-\beta_3$. Rank-one connectivity requirement reads $(\tau_{L(123)}-\tau_{2,2}):(e_1\otimes e_1)=0$. Stress field $\tau_{L(123,2)} = \beta_3\,\tau_{2,2} +(1-\beta_3)\tau_{L(123)}$ matches the effective tensor by $\tau_{L(123,2)} = \tau_0$.
\end{enumerate}

\noindent\underline{Parameters of optimal laminate}:
Rank-one connectivity requirements simplify to
\[
\ba{lll}
(\tau_{L(123)}-\tau_{2,2}):(e_1\otimes e_1)=0 &\Rightarrow &\beta_1=\dfrac{1}{(1-\beta_2)S_1}\left(\dfrac{1}{\sqrt{2}}-\beta_2\,S_1\right)\ ,\\\noalign{\smallskip}
\tau_{L(123,2)}:(e_2\otimes e_2) = \varrho &\Rightarrow &\beta_3 = \dfrac{\varrho}{\sqrt{2}\,S_2-1}
\ea
\]
and constraints on volume fractions in $Y$ are fulfilled if $|Y_1| = m_1$ and $|Y_{2,1}|+|Y_{2,2}| = m_2$, i.e.
\[
(1-\beta_2)(1-\beta_3)\beta_1 = m_1,\qquad (1-\beta_3)\beta_2 + \beta_3 = m_2 .
\]
Therefore, optimal laminate is parameterized by
\be
\label{e423-4}
\ba{ll}
\beta_1 = \dfrac{m_1}{1-m_2}, &\beta_2 = \dfrac{m_2(1-m_2)\gamma_A-m_1(m_2+\varrho)}{(1+\varrho)\big[(1-m_2)\gamma_A-m_1\big]},\\\noalign{\smallskip}
\beta_3 = \dfrac{\varrho(m_1+m_2\,\gamma_A)}{(1+\varrho-m_2)\gamma_A-m_1}, &\gamma_A = \dfrac{K_1+L_1}{K_2+L_2} .
\ea
\ee
Note that $\gamma_A\in [0,1]$.

Our next claim is that $\beta_i\in [0,1]$, $i=1,2,3$. Indeed, $\beta_1$ coincides with that in region $\cC$ and the remaining conditions can be reduced to
\be
\label{e423-5}
\ba{lcl}
\beta_2\geq 0 &\mbox{if} &m_1\leq\psi_{\cA-\cC}(m_2,\varrho),\\\noalign{\smallskip}
\beta_2\leq 1\mbox{ and }\beta_3\geq 0  &\mbox{if} &m_1\leq (1+\varrho-m_2)\gamma_A,\\\noalign{\smallskip}
\beta_3\leq 1 &\mbox{if} &m_1\leq (1-m_2)\gamma_A .
\ea
\ee
For the definition of $\psi_{\cA-\cC}(m_2,\varrho)$ and $\psi_{\cA-\cB}(m_2,\varrho)$ (used below) see Sec.\ \ref{sec321-regionBC}. It follows that the first constraint in \eqref{e423-5} is most restrictive. Hence, $\cA = \mathsf{A_1}\cup\mathsf{A_2}$ where
\[
\ba{ll}
\mathsf{A_1} = \Big\{(m_1,\varrho): &0\leq m_1\leq \psi_{\cA-\cC}(m_2,\varrho),\ \varrho\in[0,1]\Big\},\\\noalign{\smallskip}
\mathsf{A_2} = \Big\{(m_1,\varrho): &\psi_{\cA-\cC}(m_2,\varrho)\leq m_1 \leq \psi_{\cA-\cB}(m_2,\varrho),\ \varrho\in [m_2, 1]\Big\} .
\ea
\]

Laminate $L(123,2)$ (optimal in $\mathsf{A_1}$) morphs into $L(13,2)$ (optimal in $\cC$) at the boundary between regions. This follows from substituting $m_1 = \psi_{\cA-\cC}(m_2,\varrho)$ in $\beta_2$ and $\beta_3$.

\subparagraph*{Subregion $\mathsf{A_2}$\ :}
We make use of ``the coating principle", see \citep[Th.\ 9]{Alb07}, in determining optimal microstructure in subregion $\mathsf{A_2}$. Laminate $L(13_1,2,13_2)$ (optimal in region $\cB$) is coated with a layer of phase 2, in the direction $n_5 = e_1$ normal to the interface, and volume fractions $1-\beta_5$ and $\beta_5$. In this way, $L(13_1,2,13_2,2)$ is obtained. Phases 1 and 2 are thus subdivided according to $Y_1 = Y_{1,1} + Y_{1,2}$ and $Y_2 = Y_{2,1} + Y_{2,2}$ respectively. Formulae for stress fields in phases read
\be
\label{e423-7}
\ba{ll}
\tau_{1,1} = \sqrt{2}\,S_1\,e_1\otimes e_1 &\mbox{a.e.\ in } Y_{1,1},\\\noalign{\medskip}
\tau_{1,2} = \sqrt{2}\,S_1\,e_2\otimes e_2 &\mbox{a.e.\ in } Y_{1,2},\\\noalign{\smallskip}
\tau_{2,1} = \dfrac{S_2}{\sqrt{2}}(e_1\otimes e_1 + e_2\otimes e_2) &\mbox{a.e.\ in } Y_{2,1},\\\noalign{\smallskip}
\tau_{2,2} = \dfrac{S_2+f_2}{\sqrt{2}}e_1\otimes e_1 + \dfrac{S_2-f_2}{\sqrt{2}}e_2\otimes e_2 &\mbox{a.e.\ in } Y_{2,2}.
\ea
\ee

Technically, sufficient optimality condition in subregion $\cA_2$ do not restrict stress field in $Y_{2,1}$ to be spherical. However, the assumed form of $\tau_{2,1}$ proves to be optimal as it is shown in the sequel. We also assume $\tau_{2,2}:(e_1\otimes e_1)=1$ which gives $f_2=\sqrt{2}-S_2$ in \eqref{e423-7}.
\begin{figure}[H]
\centering
\includegraphics[width=90mm]{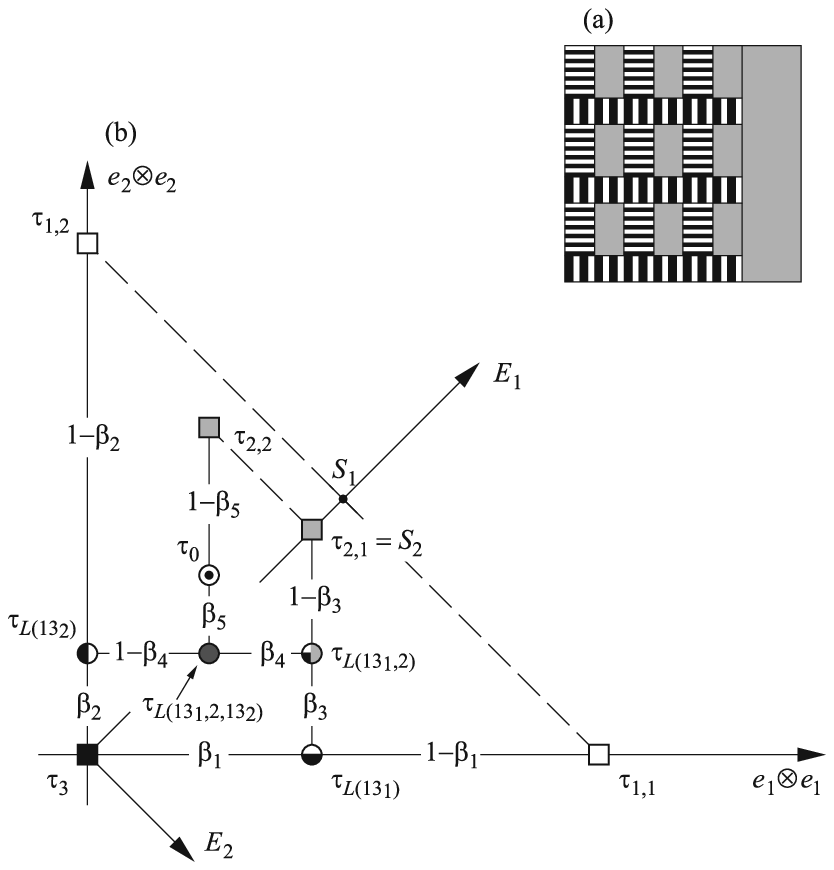}
\caption{Optimal microstructure in regions $\mathsf{A_2}$ and $\mathsf{A_2'}$: (a) layout of materials; (b) layering scheme leading to $\tau_0 = \tau_{L(13_1,2,13_2,2)}$ for region $\mathsf{A_2}$ (a mirror image with respect to $e_1\otimes e_1$ results in the scheme for region $\mathsf{A_2'}$). Stress fields in layers of strong and weak materials are represented by white and grey squares respectively, $\tau = 0$ in void is represented by black square. Circles represent stress fields in laminates.\label{fig8}}
\end{figure}

\noindent\underline{Outline of the layering scheme}:
\begin{enumerate}
\item Substructure $L(13_1,2,13_2)$ is formed along the scheme discussed in case of region $\cB$ hence it is not repeated here.

\item Final structure $L(13_1,2,13_2,2)$ is formed: second layer of phase 2 (field $\tau_{2,2}$) and $L(13_1,2,13_2)$ are laminated with $n_5 = e_1$ and volume fractions $1-\beta_5$ and $\beta_5$. Rank-one connectivity condition is given by $(\tau_{L(13_1,2,13_2)}-\tau_{2,2}):(e_1\otimes e_1)=0$. Stress field in $L(13_1,2,13_2,2)$ is linked to the effective tensor by $\tau_{L(13_1,2,13_2,2)} = \beta_5\,\tau_{2,2} +(1-\beta_5)\tau_{0,4}$ and $\tau_{L(13_1,2,13_2,2)} = \tau_0$.
\end{enumerate}

\noindent\underline{Parameters of optimal laminate}:
Rank-one compatibility of stress fields in substructures and resulting relations between laminate parameters read
\be
\label{e423-8}
\ba{lll}
(\tau_{L(13_1)}-\tau_{2,1}):(e_1\otimes e_1)=0 &\Rightarrow &\beta_1 = \dfrac{S_2}{2\,S_1}\ ,\\\noalign{\smallskip}
(\tau_{L(13_2)}-\tau_{L(13_1,2)}):(e_2\otimes e_2)=0 &\Rightarrow &\beta_2 = \dfrac{\beta_3\,S_2}{2\,S_1},\\\noalign{\smallskip}
(\tau_{L(13_1,2,13_2)}-\tau_{2,2}):(e_1\otimes e_1)=0 &\Rightarrow &\beta_4 = 1 - \dfrac{\sqrt{2}}{S_2}
\ea
\ee
and the constraints on volume fractions are given by $|Y_{1,1}|+|Y_{1,2}| = m_1$ and $|Y_{2,1}|+|Y_{2,2}| = m_2$, or explicitly
\[
\ba{l}
(1-\beta_5)\big[(1-\beta_4)(1-\beta_3)\beta_1 + \beta_4\,\beta_2\big] = m_1,\\\noalign{\smallskip}
(1-\beta_5)(1-\beta_4)\beta_3 + \beta_5 = m_2 .
\ea
\]

Introducing
\be
\label{e423-10}
x = \gamma_A\,(1+\varrho)>0,\qquad y = 2\,(m_1+m_2\,\gamma_A)>0,\qquad x\geq y
\ee
gives
\be
\label{e423-11}
\ba{l}
\beta_3 = \dfrac{x\big[(m_2+\varrho)y-2\,m_2\,x\big]}{(x-y)^2\,(1-\beta_5)},\\\noalign{\smallskip}
\beta_5 = \dfrac{m_2\,x\2-\varrho\,y\2}{(x-y)^2} .
\ea
\ee
and we proceed to show that $\beta_i\in [0,1]$, $i=1,\ldots,5$. 

From $\gamma_A\in [0,1]$ it follows that $\beta_1$ fulfills the condition. Parameter $\beta_4$ falls into required interval if $S_2\geq\sqrt{2}$ in subregion $\mathsf{A_2}$. This condition is satisfied if $2\,m_1\leq (1+\varrho-2\,m_2)\gamma_A$. Observe that
\[
m_1\leq\psi_{\cA-\cB}(m_2,\varrho)\leq \dfrac{(1+\varrho-2\,m_2)\gamma_A}{2}
\]
holds for $m_2\leq\varrho$. Thus $\beta_4\in [0,1]$.

We can assert that $\beta_5\geq 0$ if $m_2\,x\2-\varrho\,y\2\geq 0$. This inequality can be reformulated to $m_1\leq \psi_{\cA-\cB}(m_2,\varrho)$ and the assertion follows. In order to prove $\beta_5\leq 1$ we show that 
\be
\label{e423-13}
(x-y)^2 - m_2\,x\2+\varrho\,y\2\geq 0 
\ee
in entire $\mathsf{A_2}$. To this end, we first rewrite \eqref{e423-13} in the form
\be
\label{e423-14}
\ba{l}
(1+\varrho)(y-y_1)(y-y_2)\geq 0,\\\noalign{\medskip}
y_{1,2} = \left(1\pm\sqrt{(1+\varrho)m_2-\varrho}\right)\gamma_A,\qquad y_1\geq y_2,
\ea
\ee 
and the proof falls naturally into two parts. If $(1+\varrho)m_2-\varrho < 0$ then the roots in \eqref{e423-14} do not exist and \eqref{e423-13} follows immediately. 

Conversely, let us assume that $(1+\varrho)m_2-\varrho \geq 0$. Next, make use of \eqref{e423-10} and \eqref{e423-14} to calculate 
\[
m_y(m_2,\varrho) = \dfrac{1-2\,m_2+\sqrt{(1+\varrho)\,m_2-\varrho}}{2}\,\gamma_A
\]
where $m_y$ represents values of $m_1$ corresponding to $y_1$. To show that \eqref{e423-13} holds it is sufficient to check if $m_y(m_2,\varrho) < \psi_{\cA-\cC}(m_2,\varrho)$. The latter is fulfilled if $(1+\varrho)m_2-\varrho\geq -m_2$ which may be concluded from the assumption.

Comparing the expressions in \eqref{e423-11} we deduce that the discussion of $\beta_3\geq 0$ may be reduced to proving $(m_2+\varrho)y-2\,m_2\,x\geq 0$. It is straightforward to compute that this inequality is equivalent to $m_1\geq \psi_{\cA-\cC}(m_2,\varrho)$. For checking if $\beta_3\leq 1$ we write it in a form $(x-y)\big[x\,(1+m_2)-y\,(1+\varrho)\big]\geq 0$. From \eqref{e423-10} we see that it suffices to show that $x\,(1+m_2)-y\,(1+\varrho)\geq 0$. This requirement reduces to $2\,m_1\leq (1-m_2)\gamma_A $ which is valid in entire $\mathsf{A_2}$ due to
\[
m_1\leq\psi_{\cA-\cB}(m_2,\varrho)\leq \dfrac{(1-m_2)\,\gamma_A}{2} .
\]
The property of $\beta_2\in [0,1]$ follows from $\beta_3\in [0,1]$, see \eqref{e423-8}.

Recall that a boundary between regions $\mathsf{A_1}$ and $\mathsf{A_2}$ is given by $m_1 = \psi_{\cA-\cC}(m_2,\varrho)$, $\varrho\geq m_2$. By substituting this formula in $\beta_1,\ldots,\beta_5$ one may check that $L(13_1,2,13_2,2)$ (optimal in $\mathsf{A_2}$) morphs into $L(123,2)$ (optimal in $\mathsf{A_1}$). By the same token, setting $m_1 = \psi_{\cA-\cB}(m_2,\varrho)$ in $\beta_1,\ldots,\beta_5$ leads to the conclusion that $L(13_1,2,13_2,2)$ smoothly changes into $L(13_1,2,13_2)$ (optimal in $\cB$).

Similarly to $\cA$, region $\cA'$ also splits into $\mathsf{A_1}'$ and $\mathsf{A_2}'$. Optimal microstructures in both subregions are the same as in $\mathsf{A_1}$ and $\mathsf{A_2}$ respectively.

\subparagraph*{Subregion $\mathsf{A_1}'$\ :} For proving optimality of the laminate $L(123,2)$ we make use of the results obtained in Sec.\ \ref{sec324-regionA'}. Stress fields in phases read
\[
\ba{ll}
\tau_1 = \sqrt{2}\,D_{11}\,e_1\otimes e_1 &\mbox{a.e.\ in } Y_1,\\\noalign{\medskip}
\tau_{2,1} = \sqrt{2}\,D_{21}\,e_1\otimes e_1 &\mbox{a.e.\ in } Y_{2,1},\\\noalign{\smallskip}
\tau_{2,2} = \dfrac{g_1+D_{21}}{\sqrt{2}}\,e_1\otimes e_1 + \dfrac{g_1-D_{21}}{\sqrt{2}}\,e_2\otimes e_2 &\mbox{a.e.\ in } Y_{2,2}
\ea
\]
where $D_{11}$, $D_{21}$ are given by \eqref{e324-5}. We assume that $\tau_{2,2}:(e_1\otimes e_1) =1$ from which it follows that $g_1 = \sqrt{2}-D_{21}$. Further calculations are similar to those presented for subregion $\mathsf{A_1}$. They lead to formulae for optimal lamination parameters written in \eqref{e423-4} with $\varrho$ replaced by $-\varrho$.

\subparagraph*{Subregion $\mathsf{A_2}'$\ :} Laminate $L(13_1,2,13_2,2)$ proves to be optimal also in subregion $\mathsf{A_2}'$. Stress fields are given by
\[
\ba{ll}
\tau_{1,1} = \sqrt{2}\,D_{11}\,e_1\otimes e_1 &\mbox{a.e.\ in } Y_{1,1},\\\noalign{\bigskip}
\tau_{1,2} = \sqrt{2}\,D_{11}\,e_2\otimes e_2 &\mbox{a.e.\ in } Y_{1,2},\\\noalign{\smallskip}
\tau_{2,1} = \dfrac{D_{21}}{\sqrt{2}}(e_1\otimes e_1 - e_2\otimes e_2) &\mbox{a.e.\ in } Y_{2,1}, \\\noalign{\smallskip}
\tau_{2,2} = \dfrac{g_2+D_{21}}{\sqrt{2}}e_1\otimes e_1 + \dfrac{g_2-D_{21}}{\sqrt{2}}e_2\otimes e_2 &\mbox{a.e.\ in } Y_{2,2}
\ea
\]
and we assume that $\tau_{2,2}:(e_1\otimes e_1) = 1$ hence $g_2=\sqrt{2}-D_{21}$. Optimal lamination parameters are derived similarly to those in region $\mathsf{A_2}$. They read 
\[
\ba{ll}
\beta_1 = \dfrac{D_{21}}{2\,D_{11}}, &\beta_2 = \dfrac{D_{21}\,\beta_3}{2\,D_{11}},\\\noalign{\medskip}
\beta_3 = \dfrac{x'\,[(m_2-\varrho)y'-2\,m_2\,x']}{(x'-y')^2(1-\beta_5)}, &\beta_4 = 1-\dfrac{\sqrt{2}}{D_{21}},\\\noalign{\medskip}
\beta_5 = \dfrac{m_2\,(x')^2+\varrho(y')^2}{(x'-y')^2}
\ea
\]
where $x' = (1-\varrho)\,\gamma_A$ and $y'=y$.

Proof of $\beta_i\in [0,1]$, $i=1,\ldots,5$ in subregion $\mathsf{A_2}'$ follows the pattern set in the discussion regarding $\mathsf{A_2}$.

\subsection{Alternative optimal structures}
\label{sec40-altstruct}

\begin{figure}[H]
\centering
\includegraphics[width=90mm]{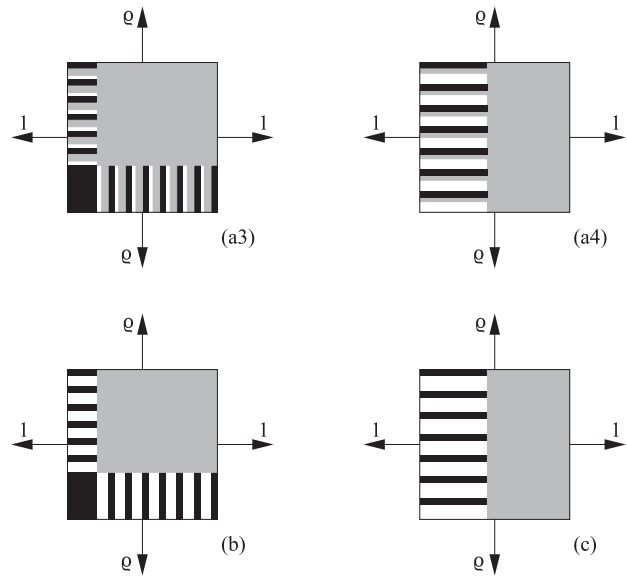}
\caption{ Optimal Sigmund-Gibiansky-type structures and their high-rank laminate limits: (a3) SG-type structure in subregion $\mathsf{A_3}$; (a4) rank-2 laminate in subregion $\mathsf{A_4}$; (b) SG-type structure in region $\cB$; (c) rank-2 laminate (``T-structure'') in region $\cC$. Regions of optimality are shown in Fig.\ \ref{fig10}.\label{fig9}}
\end{figure}

\subparagraph*{Structures of Sigmund and Gibiansky}
Sufficient optimality conditions set the requirements for stress fields in materials within optimal structures, but not for the parameters of optimal geometries. Here, we describe an alternative class of optimal structures inspired by the approach of Sigmund \& Gibiansky \citep{Sig00, Gib00} and we show that their results can be generalized beyond the isotropic case\footnote{The authors are indebted to the anonymous reviewer for bringing this issue to their attention.}. Layouts of materials in Sigmund-Gibiansky-type (SG-type) structures and their high-rank laminate limits are shown in Fig.\ \ref{fig9}.
\begin{figure}[H]
\centering
\includegraphics[width=90mm]{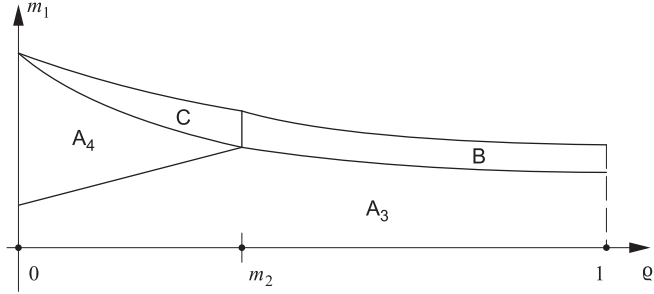}
\caption{High-porosity regions of optimality of structures from Fig.\ \ref{fig9}. Region $\cB$ and $\cC$ coincide with those shown in Fig.\ \ref{fig2}. The property of subregions $\mathsf{A_1}$, $\mathsf{A_2}$ in Fig.\ \ref{fig2} and $\mathsf{A_3}$, $\mathsf{A_4}$ above is that $\mathsf{A_3}\cup\mathsf{A_4} = \mathsf{A_1}\cup\mathsf{A_2} = \cA$.\label{fig10}}
\end{figure}

Consider region $\cB$. From sufficient optimality conditions, see Sec.\ \ref {sec422-regionBB'}, we know that optimal stresses in layers of phase 1 are unidirectional, $ \det \tau(y) = 0, y\in Y_1$. This condition is satisfied if material 1 is laminated with void; the density of the field in the $L(13)$ laminate is constant everywhere, as is the density of the stress inside layers of phase 1. The value of $\det\tau(y)$ in inner points of $Y_1$ tends to zero when the thickness-to-length ratio of layers decreases. The stress tensor in phase 2 is spherical, $\tau(y)\sim I, y\in Y_2$ and $I$ stands for a second-rank unit tensor. High-rank laminate obeying the mentioned conditions is shown in Fig.\ \ref{fig6}, we show that they are also satisfied in a SG-type structure from Fig.\ \ref{fig9}(b).
\begin{figure}[H]
\centering
\includegraphics[width=90mm]{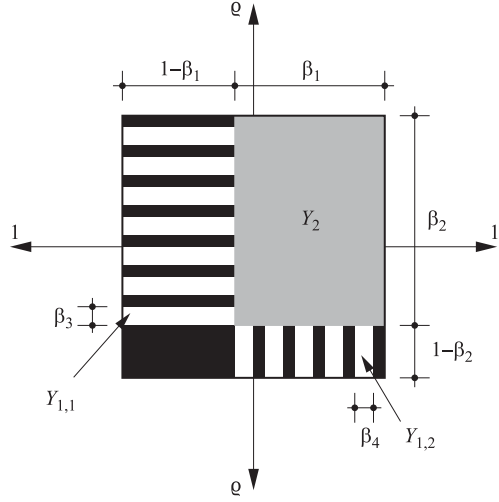}
\caption{Loading and geometry of the Sigmund-Gibiansky-type structure optimal in region $\cB$. Symbols $Y_{1,1}$, $\beta_3$ and $Y_{1,2}$, $\beta_4$ refer to the total fraction of phase 1 and its overall thickness in perpendicularly oriented $L(13)$ laminates.\label{fig11}}
\end{figure}

Assume that a square cell of periodicity is divided into four rectangles. Two opposite rectangles are filled with pure phases 2 or 3 (void), and two remaining ones are filled with $L(13)$ laminates transferring the load towards phase 2. Geometric parameters of the structure are explained in Fig.\ \ref{fig11}. They are related to the volume fractions of non-degenerate phases as following
\be
\label{eq40-1}
\ba{ll}
m_1 = (1-\beta_1)\beta_2\beta_3 + (1-\beta_2)\beta_1\beta_4, &m_2 = \beta_1\beta_2,\\\noalign{\smallskip}
\beta_i\in [0, 1], i=1,\ldots,4 .
\ea
\ee
If the external field $\tau_0 = 1\,e_1\otimes e_1 + \varrho\,e_2\otimes e_2$, $\varrho\in [0, 1]$, is applied to such a structure then optimal fields in materials are given by
\be
\label{eq40-2}
\ba{ll}
\tau_{1,1} = \dfrac{1}{\beta_2\beta_3}\,e_1\otimes e_1 &\mbox{a.e. in }Y_{1,1}, \\\noalign{\smallskip}
\tau_{1,2} = \dfrac{\varrho}{\beta_1\beta_4}\,e_2\otimes e_2 &\mbox{a.e. in }Y_{1,2}, \\\noalign{\smallskip}
\tau_2 = \dfrac{1}{\beta_2}\,e_1\otimes e_1 +  \dfrac{\varrho}{\beta_1}\,e_2\otimes e_2 &\mbox{a.e. in }Y_2, \\\noalign{\smallskip}
\ea
\ee
where
\be
\label{eq40-3}
\dfrac{1}{\beta_2\beta_3} = \dfrac{\varrho}{\beta_1\beta_4}, \qquad \dfrac{1}{\beta_2} = \dfrac{\varrho}{\beta_1} .
\ee
From \eqref{eq40-1} it follows that
\be
\label{eq40-4}
\ba{lll}
\beta_1 = \sqrt{\varrho\,m_2}, &\beta_2 = \sqrt{\dfrac{m_2}{\varrho}}, &\beta_3=\beta_4 = \dfrac{\varrho\,m_1}{(1+\varrho-2\,\sqrt{\varrho\,m_2})\,\sqrt{\varrho\,m_2}} .
\ea
\ee
Substituting \eqref{eq40-4} in \eqref{eq40-2} gives \eqref{e422-1}. Consequently, one may conclude that the anisotropic SG-type structures from Fig.\ \ref{fig9}(b) are optimal in entire region $\cB$.

The result obtained above has a clear physical interpretation. Stress field in phase 2 is isotropic, but the rectangle $Y_2$ is elongated against the larger component of average stress so that an uneven loading is supported. When the elongation reaches its limit, $\beta_2=1$, the structure is transformed into a ``T-structure'' shown in Fig.\ \ref{fig9}(c) that is optimal in region $\cC$, see Fig.\ \ref{fig5}. 

Similar considerations prove optimality of the SG-type structures from Fig.\ \ref{fig9}(a3) in the subregion $\mathsf{A_3}$ in Fig.\ \ref{fig10}. The elongation of the rectangular domain containing bulk portion of phase 2 reaches its limit on the boundary with $\mathsf{A_4}$. After this, the structure is transformed into a rank-2 laminate from Fig.\ \ref{fig9}(a4) that is optimal in region $\mathsf{A_4}$, see also Fig.\ \ref{fig7}. Details of calculations are similar to the above.

\emph{Remark 1:}\\ 
Sufficient optimality conditions are the same in whole region $\cA$; they are realized by different structures in different subregions. The division of $\cA$ into subregions $\mathsf{A_3}$, $\mathsf{A_4}$ in Fig.\ \ref{fig10} does not coincide with regions $\mathsf{A_1}$, $\mathsf{A_2}$ in Fig.\ \ref{fig2}. This is due to the additional assumption of the stress field isotropy in the rectangle of phase 2, see Fig.\ \ref{fig9}(a3).

\emph{Remark 2:}\\
It is truly remarkable that isotropic structures in region $\cA$ were correctly predicted in the pioneering publication by \citet{Gib00} in the absence of sufficient optimality conditions found in the present paper. 

Now, with the systematic use of these conditions, we also demonstrate the optimal SG-type structures for region $\cB$ thus improving the intuitive results of \citep{Gib00}.

\subparagraph*{Number of length scales in optimal microstructures}
Optimal two-material composites can take a form of single-scale Vigdergauz structure in which weak material is embedded in the strong one, see \citep{Vig89}. Composites considered in the present paper require at least two scales. Indeed, sufficient optimality conditions in regions $\cA$, $\cB$ and $\cC$ state that the stress tensor in phase 1 is unidirectional, $\det\tau=0$. This in turn means that phase 1 must be laminated with a void in a smaller scale; in this case $\det\tau\to 0$ everywhere in phase 1.

%% file: 3mat-acgd-revised-sec5.tex
The detailed description of the mentioned regions $\cD$, $\cD'$ and $\cE$ of large volume fraction of the first material, or, equivalently, low-porosity regions, will be provided in a separate paper. Here we restrict ourselves with some brief remarks outlining the current results. 

\subparagraph*{Region $\mathsf{D}$: The Hashin-Shtrikman bound on bulk modulus}
At the boundary of regions $\cB$ and $\cC$ that correspond to maximal allowed volume fraction $m_1$, the optimal translation parameter reaches the value of $L_1$. The energy bound $U_\mrm{tr}(\varrho)$ in region $\cB$ transforms into the classical translation bound which corresponds to the Hashin - Shtrikman bound on the bulk modulus for isotropic composites. This bound is realizable, see \citep{Gib00,Che09,Che12}. The anisotropic translation bound is attained on certain microstructures only when the anisotropy level is not too large, compare the discussion in \citep{Che11}. The optimal structures for both conducting and elastic composites are similar, they are determined by high-rank orthogonal laminates $L(13_1,2,13_2,1,1)$. These structures are obtained by enveloping the nucleus laminate $L(13_1,2,13_2)$ - optimal for the region $\cB$ - by two orthogonal layers of the first material. It is shown in \citep{Alb07} that such enveloping is stable with respect to the translation bound: if the nucleus satisfies this bound, then the enveloped nucleus also satisfies it.

\subparagraph*{Region $\mathsf{D'}$} Similarly, at the boundary of regions $\cB'$ and $\cC'$, the optimal translation parameter reaches the value of $-K_1$. However, in this case, the energy bound $U_\mrm{tr}(\varrho)$ in region $\cB'$ \emph{does not} give rise to the Hashin - Shtrikman bound on the shear modulus for isotropic composites. 

Indeed, $U_{\rm{tr}}$ measures the energy of a~composite subjected to an arbitrary stress field whose anisotropy is controlled by $\varrho\in [-1, 1]$. Consequently, if we set $\varrho=1$ then the effective energy is optimized only in a direction of the applied field $\tau_0=[(1+\varrho)/2]\,E_1$ which is spherical, i.e.\ isotropic.

On the contrary, setting $\varrho=-1$ does not lead to a~similar conclusion because applying the deviatoric field $\tau_0=[(1-\varrho)/2]\,E_2$ and retaining the isotropy of a~composite medium by controlling its response in the direction $E_3$ at the same time is impossible.

\subparagraph*{Region $\cE$: Guessed optimal structures}
Optimal  $L(13_1,2,13_2,1,1)$ structures degenerate into the most anisotropic $L(13, 2, 1)$, when the ani\-so\-tropy of the external field increases (the value of $\varrho$ decreases from 1 towards 0). When the anisotropy level increases even further, the translation bound is not realizable by the known structures. Moreover, it is definitely not optimal for strongly anisotropic structures; the reasons are discussed in \citep{Che11}.  

We conjecture that the region $\cE$ of large volume fractions $m_1$ and strongly anisotropic loadings correspond to the limiting structures $L(13,2,1)$. The bound for this region is presently unknown, and we guess that it corresponds to another inequality that becomes an equality in that region. To support our guess we mention that: 
\begin{itemize}
\item[-] the optimal structure in Region $\cC$ is $L(13,2)$. The $L(13,2,1)$ structures degenerate into them, when the fraction of external layer of the first material vanishes;
\item[-] the best known bounds for extremely anisotropic structures ($\varrho \to 0$) correspond to the same structure $L(13,2,1)$ in that region, see \citep{Che96};
\item[-] the structures that realize the translation bound for moderately ani\-so\-tro\-pic loadings, also degenerate into $L(13,2,1)$;
\item[-] the $L(13,2,1)$ structures degenerate into $L(13)$ when the fraction of the second material disappears;
\item[-] the $L(13,2,1)$ structures degenerate into $L(12)$ when the fraction of the third material (void) disappears.
\end{itemize}

In the absence of the bound, one cannot prove the sufficient optimality conditions for the guessed structures and therefore the global character of their optimality. It can be numerically shown, however, that the relative gap between a rough bound for the energy and a structure of this class is very small, see \citep{Che11}. Therefore, these structures are either optimal or a close approximation of optimal, and can be treated as optimal for practical purposes. 